\documentclass[a4paper,11pt]{article}
\pdfoutput=1 

\usepackage{jinstpub} 

\usepackage{hyperref}
\usepackage{siunitx}
\usepackage{amsmath}
\usepackage{bm}
\usepackage{caption}
\usepackage{subcaption}
\usepackage{float}
\usepackage{graphicx}
\usepackage[siunitx,oldvoltagedirection]{circuitikz}
\usetikzlibrary{decorations.markings}

\title{\boldmath Neural network--featured timing systems for radiation detectors: performance evaluation based on bound analysis}

\author[a,b]{Pengcheng Ai,}
\author[a,b,1]{Zhi Deng,\note{Corresponding author.}}
\author[a,b]{Yi Wang}
\author[a,b]{and Linmao Li}

\affiliation[a]{Department of Engineering Physics, Tsinghua University,\\Haidian District, Beijing, 100084, P. R. China}
\affiliation[b]{Key Laboratory of Particle and Radiation Imaging (Tsinghua University), Ministry of Education,\\Haidian District, Beijing, 100084, P. R. China}

\emailAdd{dengz@mail.tsinghua.edu.cn}

\abstract{Waveform sampling systems are used pervasively in the design of front end electronics for radiation detection. The introduction of new feature extraction algorithms (eg. neural networks) to waveform sampling has the great potential to substantially improve the performance and enrich the capability. To analyze the limits of such algorithms and thus illuminate the direction of resolution optimization, in this paper we systematically simulate the detection procedure of contemporary radiation detectors with an emphasis on pulse timing. Neural networks and variants of constant fraction discrimination are studied in a wide range of analog channel frequency and noise level. Furthermore, we propose an estimation of multivariate Cram\'er Rao lower bound within the model using intrinsic-extrinsic parametrization and prior information. Two case studies (single photon detection and shashlik-type calorimeter) verify the reliability of the proposed method and show it works as a useful guideline when assessing the abilities of various feature extraction algorithms.}

\keywords{Analysis and statistical methods; Pattern recognition, cluster finding, calibration and fitting methods; Timing detectors; Electronic detector readout concepts (solid-state)}

\arxivnumber{2105.14687} 

\begin{document}
\maketitle
\flushbottom

\section{Introduction}

Waveform sampling systems based on modern analog-to-digital converters (ADC)
are in the research \& development for applications of radiation detection, such as high energy physics \cite{CERN-LHCC-2017-011,Atanov:2018ich,Semenov_2020} and positron emission tomography (PET) \cite{Han_2019,PARK2019117}. ADCs with sufficient precision and sampling rates are able to preserve the details of the original signal and thus provide flexibility when multiple aspects (time, energy, position) of physical information are desired. Another advantage of waveform sampling systems lies in their guaranteed resolution when configured and operated in proper conditions \cite{FALLULABRUYERE2007247}. Usually sub-\si{ns} timing resolution could be achieved by an ADC with a sampling rate of $\sim$100 \si{MS/s}.

In the past decade, neural networks with crafted deep structures (aka. deep learning) \cite{8694781} drew much attention in the community and were applied to multiple disciplines. For example, in high energy physics, deep learning has been proposed for or used in many tasks from high-level physics to low-level detection \cite{doi:10.1146/annurev-nucl-101917-021019}. In the development of front-end electronics for particle detectors, neural networks have been applied to pulse characterization \cite{Ai_2019,AI2020164420} and exhibited superior performance compared to curve fitting. These results, in combination with waveform sampling techniques, show a new direction to reform and intelligentize the existing front-end systems.

Recent studies on neural networks in radiation detection focused on demonstrating their effectiveness with high-speed digitizers \cite{Berg_2018,8844693}. Although inspiring results have been showed, the limitation and potential of these waveform-based methods are still in the mist. For one thing, we would like to know the limits of achievable resolution when we acquire information from sampled waveform; for another, we would like to know how much room for improvement of performance with more advanced algorithms. With these understandings (especially the latter one), we can guide the development of future data acquisition systems to meet the specification with minimum cost. This is particularly true of neural networks when we always need to trade-off between accuracy and computational complexity when putting them into practice.

In this paper, we aim to provide understandings of waveform sampling systems by simulating the detection procedure based on scintillator, silicon photomultipliers (SiPM) and feature extraction algorithms. For the front end, we investigate two different settings of radiation detection, i.e. single photon detection and shashlik-type calorimeter (a kind of sampling calorimeter with alternative scintillators and absorbers). For the back end, we apply traditional constant fraction discrimination (CFD) as well as novel neural networks to digital samples. Besides, a modelling technique based on parametrization of waveform variations is proposed for bound analysis when waveform characteristics are related to certain physical information. This modelling technique is applicable to a wide variety of feature extraction tasks, but we will center on the pulse timing problem as typical applications.

The main contributions of the paper are listed below:

\begin{itemize}
	\item We construct a common framework to study timing systems based on waveform sampling and subsequent feature extraction. This framework is compatible with both the traditional timing methods and emerging intelligent algorithms.
	
	\item We propose a new modelling method directly on waveform sampling with \emph{double-domain} parametrization and \emph{prior} information. A derivation process utilizing the above model is provided to compute the lower bound of variance when extracting features from the waveform.
	
	\item Two applications (single photon detection and shashlik-type calorimeter) are investigated in simulation studies. Theoretical estimations are compared to several timing algorithms to demonstrate the advantage of neural networks and validate the practicality of the modelling method.
\end{itemize}

\section{Related work}

Classical feature extraction takes advantage of \emph{ad hoc} algorithms for each target, for example, trapezoidal filtering or charge integration for energy measurement, and leading edge or constant fraction discrimination for time measurement. In contrast, well-structured neural networks are universal approximators \cite{ZHOU2020787,10.1007/11840817_66}, and they have the potential to fulfill multiple tasks with similar architectures. From the perspective of neural networks, they "see" sampling points of waveform and "produce" continuous outputs for certain targets. They should exploit information residing in the discrete time series to achieve the goal of predicting some physics-related information.

In high energy physics, abundant and diversified data provide a great opportunity for neural networks to take part in many low-level to medium-level tasks. For example, \cite{Belayneh:2019vyx,Ai_2018} use three-dimensional convolutional neural networks (CNN) for particle/event identification and energy regression in a sensitive volume. \cite{PhysRevD.99.012011,AI2020164640} utilize two-dimensional CNNs for regression of particular physical information (position, energy) on a grid. Besides, one-dimensional CNNs have been applied to pulse timing for upgrades of calorimeters in ALICE experiment \cite{Ai_2019}. Digital logic of the neural network accelerator has been implemented in \cite{Chen2020,AI2020164420}.

To assess the performance of neural networks, it is helpful to know their limits, or lower bound, to estimate the room for improvement by adjusting architectures. Previous work \cite{CLINTHORNE1990157,Seifert_2012,Cates_2015,GUNDACKER20156,8049484} about lower bound of scintillation detectors used the Cram\'er Rao Bound theory \cite{Cramer:107581} and derived their equations considering the detecting principle of the detector. Among them, \cite{CLINTHORNE1990157,Seifert_2012} studied crystals coupled with photomultipliers, while \cite{Seifert_2012,Cates_2015,GUNDACKER20156,8049484} introduced the SiPMs which are more compact and accurate in timing. Relatively, less work in the literature discussed the lower bound of waveform sampling in the context of nuclear physics. \cite{MENG2005435} analyzed the timing resolution of CZT detectors (scintillation detectors made by CdZnTe) and computed the Cram\'er Rao Bound with the approximated covariance matrix of sampling points. However, they treated all parameters of the waveform in a uniform manner (single-domain parametrization) and did not take the prior distribution of parameters into account.

\section{Methodology and simulation setup}

\begin{figure*}[htb]
	\centering
	\includegraphics[width=0.85\linewidth]{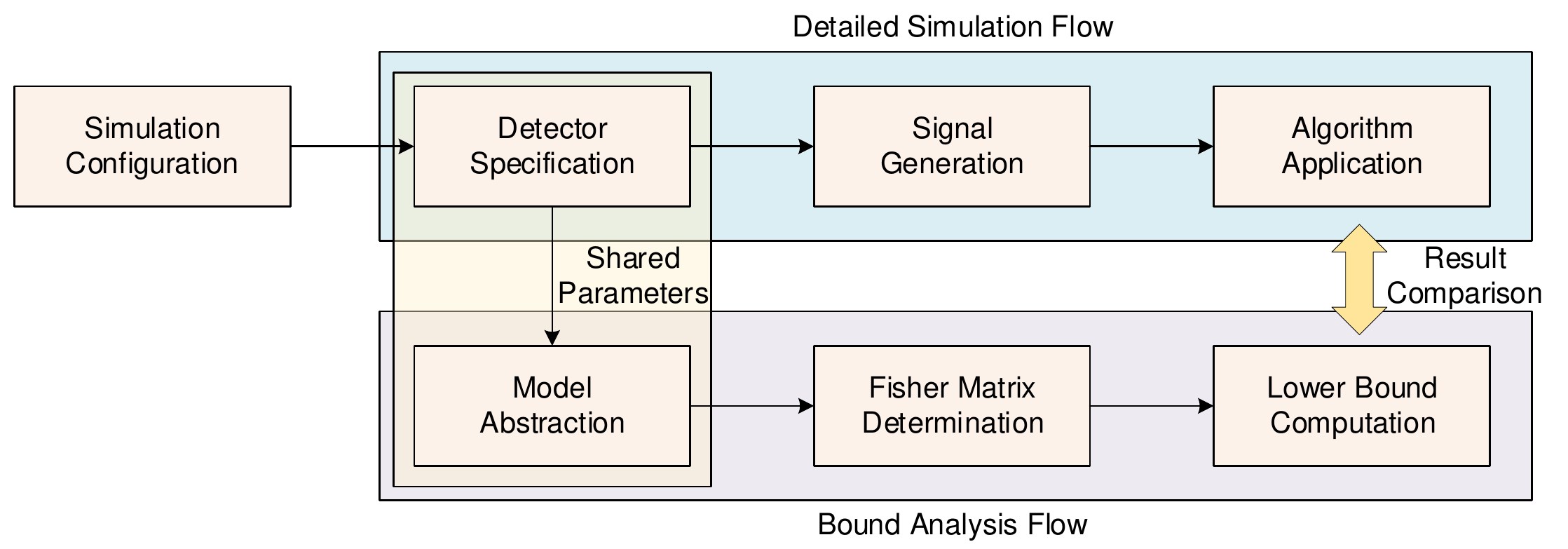}
	\caption{The proposed methodology with two parallel flows conducted after the parameters of the detector are specified. The results from detailed simulation and bound analysis are compared for better understanding of available timing algorithms.}
	\label{fig:methodology}
\end{figure*}

To study the radiation detectors based on waveform sampling and analyze the bound of various timing algorithms in a consistent manner, a unified framework for this research is proposed, shown in figure \ref{fig:methodology}. The overall architecture and principle of the detector are consolidated in the beginning phase (\emph{Simulation Configuration}). Then each part of the detector is specified with known parameters (\emph{Detector Specification}) which are shared across the \emph{Detailed Simulation Flow} and the \emph{Bound Analysis Flow}. For the upper part, stimulus from the source event is propagated through the whole detector flow to generate waveform sampling points (\emph{Signal Generation}). Afterwards, several timing algorithms are applied to the sampling data for accurate time measurement (\emph{Algorithm Application}). For the lower part, \emph{Model Abstraction} transforms the detector flow into a mathematical model, and the information matrix is calculated in \emph{Fisher Matrix Determination}. Finally, \emph{Lower Bound Computation} estimates the limit of achievable timing resolution, and results from two flows are compared together. The detailed mathematical deduction of the lower bound can be found in section \ref{sec:CRLB_for_waveform}.

\begin{figure*}[htb]
	\centering
	\includegraphics[width=0.95\linewidth]{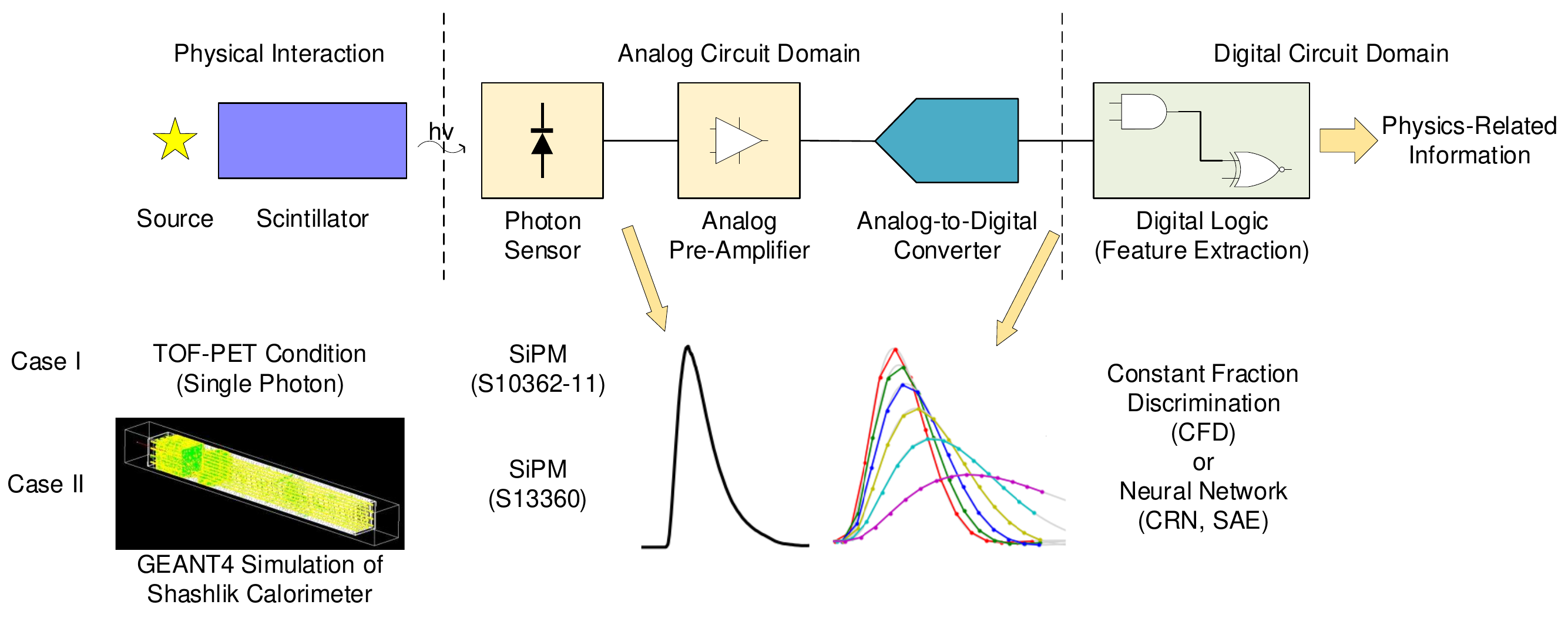}
	\caption{The radiation detection procedure based on scintillator, SiPM and waveform sampling back end.}
	\label{fig:detector_flow}
\end{figure*}

The waveform sampling--based detection procedure for this research is shown in figure \ref{fig:detector_flow}. This procedure is a common setup for simulation as well as experiments in many scenarios of radiation detection.

At the front end, the electrical signal produced by the photon sensor originates from scintillation light of an incident particle. Primarily, we consider two cases: single photon signal of SiPM and cumulative signal of SiPM. The former one is heavily researched (in the form of single photon time resolution, or SPTR) to judge the timing performance of SiPMs for medical imaging (such as time-of-flight PET) and other applications. The reference time in this case is the starting point when an avalanche is trigger in a single cell of the SiPM.

For the cumulative signal, we conduct physical simulation of a shashlik-type calorimeter with the \textsc{Geant4} simulation toolkit \cite{AGOSTINELLI2003250}. The single tower of the detector is a sampling calorimeter with alternative polystyrene scintillators and lead plates \cite{Semenov_2020}. 16 wavelength shifting fibers penetrate the bulk and guide the photons onto a SiPM at one side. A sketch of the tower structure is shown at the left bottom corner of figure \ref{fig:detector_flow}. The detailed parameters of the detector geometry can be found in \cite{NICA-MPD-ECAL-TDR}. The 1 GeV electron is chosen as the incident particle in the physical simulation. The reference time in this case is the time origin when the incident particle moves from its initial location.

The analog signals in above two cases have different statistical characteristics, so we need case-specific analysis for them. The transient response of SiPM devices in both cases is described in section \ref{sec:modelling_sipm}. Besides, we elaborate on the modelling of single photon signal and cumulative signal of SiPMs and their associated lower bounds in section \ref{sec:case_study_LB}.

At the back end, ADC converts the analog signal into a digital time series, which is ready for feature extraction implemented by digital logic. Two categories of timing algorithms are studied: variants of constant fraction discrimination (CFD) and neural networks. We will introduce these timing algorithms in section \ref{sec:overview_timing_algo}. Currently we implement them through computer software, and transforming them into hardware on application specific integrated circuits or field programmable gate arrays is straightforward.

\section{Modelling of SiPM devices}
\label{sec:modelling_sipm}

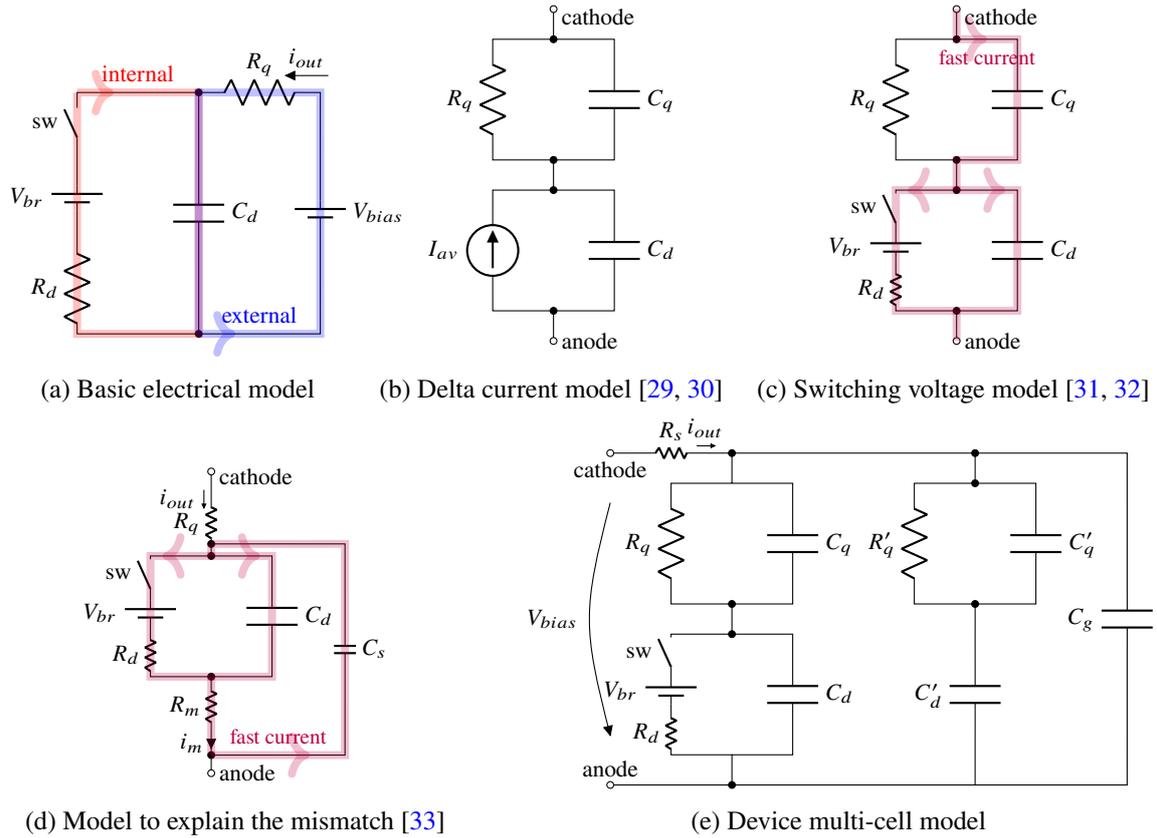
\begin{figure*}[htb]
	\centering
	\begin{subfigure}[b]{0.3\textwidth}
		\centering
		\begin{circuitikz}[scale = 0.8, transform shape] \draw
			(0, 0) to[R, l=$R_d$] (0, 1.5)
			to[battery1, l=$V_{br}$, invert] (0, 3)
			to[nos, l=sw] (0, 4) -- (2, 4)
			to[C, l=$C_d$, *-*] (2, 0) -- (0, 0)
			(2, 4) to[R, l=$R_q$, f<=$i_{out}$] (4, 4)
			to[battery1, l=$V_{bias}$] (4, 0) -- (2, 0)
			;
			\begin{scope}[very thick,decoration = {
					markings,
					mark = at position 0.05 with {\arrow{>}}}
				] 
				\draw[line width = 3pt, red, opacity = 0.25, postaction = {decorate}]
				(0, 4) -- (2, 4) node[above, midway, opacity = 1]{internal}
				-- (2, 0) -- (0, 0) -- (0, 4)
				;
			\end{scope}
			\begin{scope}[very thick,decoration = {
					markings,
					mark = at position 0.05 with {\arrow{>}}}
				] 
				\draw[line width = 3pt, blue, opacity = 0.25, postaction = {decorate}]
				(2, 0) -- (4, 0) node[above, midway, opacity = 1]{external} -- (4, 4) -- (2, 4) -- (2, 0)
				;
			\end{scope}
		\end{circuitikz}
		\caption{Basic electrical model}
		\label{fig:basic_elec_model}
	\end{subfigure}
	\hfill
	\begin{subfigure}[b]{0.3\textwidth}
		\centering
		\begin{circuitikz}[scale = 0.8, transform shape] \draw
			(0, 0) node[right]{anode} to[short, o-*] (0, 0.5) -- (-1, 0.5)
			to[american current source, l=$I_{av}$] (-1, 2.5) -- (1, 2.5) 
			to[C, l=$C_d$] (1, 0.5) -- (0, 0.5)
			(0, 2.5) to[short, *-*] (0, 3) -- (-1, 3)
			to[R, l=$R_q$] (-1, 5) -- (1, 5)
			to[C, l=$C_q$] (1, 3) -- (0, 3)
			(0, 5) to[short, *-o] node[right]{cathode} (0, 5.5)
			;
		\end{circuitikz}
		\caption{Delta current model \cite{4179230,CORSI2007416}}
		\label{fig:delta_current_model}
	\end{subfigure}
	\hfill
	\begin{subfigure}[b]{0.35\textwidth}
		\centering
		\begin{circuitikz}[scale = 0.8, transform shape] \draw
			(0, 0) node[right]{anode} to[short, o-*] (0, 0.5) -- (-1, 0.5)
			to[R, l=$R_d$, bipoles/length=0.6cm] (-1, 1.2)
			to[battery1, l=$V_{br}$, invert] (-1, 1.9)
			to[nos, l=sw] (-1, 2.5) -- (1, 2.5) 
			to[C, l=$C_d$] (1, 0.5) -- (0, 0.5)
			(0, 2.5) to[short, *-*] (0, 3) -- (-1, 3)
			to[R, l=$R_q$] (-1, 5) -- (1, 5)
			to[C, l=$C_q$] (1, 3) -- (0, 3)
			(0, 5) to[short, *-o] node[right]{cathode} (0, 5.5)
			;
			\begin{scope}[very thick,decoration = {
					markings,
					mark = at position 0.2 with {\arrow{>}}}
				] 
				\draw[line width = 3pt, purple, opacity = 0.25, postaction = {decorate}]
				(0, 5.5) -- (0, 5) -- node[below, midway, opacity = 1]{\small fast current} (1, 5) -- 
				(1, 3) -- (0, 3) -- (0, 2.5)
				;
				\draw[line width = 3pt, purple, opacity = 0.25, postaction = {decorate}] (0, 2.5) -- (1, 2.5) -- (1, 0.5) -- (0, 0.5)
				;
				\draw[line width = 3pt, purple, opacity = 0.25, postaction = {decorate}] (0, 2.5) -- (-1, 2.5) -- (-1, 0.5) -- (0, 0.5)
				;
				\draw[line width = 3pt, purple, opacity = 0.25] (0, 0.5) -- (0, 0)
				; 
			\end{scope}
		\end{circuitikz}
		\caption{Switching voltage model \cite{5341428,ACERBI201916}}
		\label{fig:switching_vol_model}
	\end{subfigure}
	\hfill
	\begin{subfigure}[b]{0.4\textwidth}
		\centering
		\begin{circuitikz}[scale = 0.8, transform shape] \draw
			(0, -1.1) node[right]{anode} to[short, o-] (0, -0.7)
			to[short, i<=$i_m$] (0, -0.5)
			to[R, l=$R_m$, bipoles/length=0.6cm] (0, 0.5)
			to[short, -*] (0, 0.5)
			(0, 0.5) -- (-1, 0.5)
			to[R, l=$R_d$, bipoles/length=0.6cm] (-1, 1.2)
			to[battery1, l=$V_{br}$, invert] (-1, 1.9)
			to[nos, l=sw] (-1, 2.5) -- (1, 2.5) 
			to[C, l=$C_d$] (1, 0.5) -- (0, 0.5)
			(0, 2.5)
			to[short, *-] (0, 2.5)
			to[R, l=$R_q$, bipoles/length=0.6cm, f<=$i_{out}$] (0, 3.6)
			to[short, -o] node[right]{cathode} (0, 3.9)
			(0, 2.7) to[short, *-] (2.2, 2.7)
			to[C, l=$C_s$, bipoles/length=0.6cm] (2.2, -0.8) to[short, -*] (0, -0.8)
			;
			
			\begin{scope}[very thick,decoration = {
					markings,
					mark = at position 0.2 with {\arrow{>}}}
				] 
				\draw[line width = 3pt, purple, opacity = 0.25, postaction = {decorate}]
				(0, -0.8) -- node[above, midway, opacity = 1]{\small fast current} (2.2, -0.8) -- (2.2, 2.7) -- (0, 2.7)
				(0, 2.7) -- (0, 2.5)
				;
				\draw[line width = 3pt, purple, opacity = 0.25, postaction = {decorate}] (0, 2.5) -- (1, 2.5) -- (1, 0.5) -- (0, 0.5)
				;
				\draw[line width = 3pt, purple, opacity = 0.25, postaction = {decorate}] (0, 2.5) -- (-1, 2.5) -- (-1, 0.5) -- (0, 0.5)
				;
				\draw[line width = 3pt, purple, opacity = 0.25] (0, 0.5) -- (0, -0.8)
				; 
			\end{scope}
		\end{circuitikz}
		\caption{Model to explain the mismatch \cite{Cova:96}}
		\label{fig:model_exp_mismatch}
	\end{subfigure}
	\hfill
	\begin{subfigure}[b]{0.55\textwidth}
		\centering
		\begin{circuitikz}[scale = 0.8, transform shape] \draw
			(0, 0) -- (0, 0.5) -- (-1, 0.5)
			to[R, l=$R_d$, bipoles/length=0.6cm] (-1, 1.2)
			to[battery1, l=$V_{br}$, invert] (-1, 1.9)
			to[nos, l=sw] (-1, 2.5) -- (1, 2.5) 
			to[C, l=$C_d$] (1, 0.5) -- (0, 0.5)
			(0, 2.5) to[short, *-*] (0, 3) -- (-1, 3)
			to[R, l=$R_q$] (-1, 5) -- (1, 5)
			to[C, l=$C_q$] (1, 3) -- (0, 3)
			(0, 5) -- (0, 5.5)
			(0, 0) -- (4, 0) -- (4, 0.5)
			to[C, l=$C_d^{\prime}$] (4, 2.5) to[short, -*] (4, 3) -- (3, 3)
			to[R, l=$R_q^{\prime}$] (3, 5) -- (5, 5)
			to[C, l=$C_q^{\prime}$] (5, 3) -- (4, 3)
			(4, 5) to[short, *-*] (4, 5.5) -- (0, 5.5)
			(4, 0) -- (6.5, 0)
			to[C, l=$C_g$] (6.5, 5.5) -- (4, 5.5)
			(-2, 0) node[above]{anode} to[short, o-*] (0, 0)
			(-2, 5.5) node[below]{cathode} to[short, o-] (-1.9, 5.5)
			to[R, l=$R_s$, f=$i_{out}$, bipoles/length=0.6cm] (-0.1, 5.5) to[short, -*] (0, 5.5)
			(-2, 0) to[open, v^<=$V_{bias}$] (-2, 5.5)
			;
		\end{circuitikz}
		\caption{Device multi-cell model}
		\label{fig:device_multi_model}
	\end{subfigure}
	\caption{Several electrical models of the single-cell circuits in SiPM, and a whole device model.}
	\label{fig:several_elec_single_cell}
\end{figure*}

SiPM devices are shown to have good timing abilities so as to be applied in upgrades of radiation detection instruments. Figure \ref{fig:several_elec_single_cell} shows several electrical models of the single-cell circuit, aka. the single-photon avalanche diode (SPAD), and also a multi-cell model for the whole device.

Figure \ref{fig:basic_elec_model} is the basic electrical model only considering the quenching resistance $R_q$ in the external circuit of SPAD. The avalanche is simulated as a switch $sw$ and a voltage source $V_{br}$. The diode resistance $R_d$ is so small that the leading edge of current pulse is dominated by the time constant $R_d \cdot C_d$. After quenching, the current pulse declines with the time constant $R_q \cdot C_d$.

Figure \ref{fig:delta_current_model} further considers the capacitance $C_q$ in parallel with $R_q$, and the avalanche is regarded as a delta current source (a pulse with infinitely small duration and certain integral value) with predefined charge injection \cite{4179230,CORSI2007416}. This model does not describe the leading edge of the current pulse while the trailing edge is more accurate.

Figure \ref{fig:switching_vol_model} combines the above two models to cover both the leading edge and the trailing edge with a switching voltage source. It should be noted that the time constant for the leading edge is no longer determined by $\tau_r = (C_d + C_q) \cdot (R_d || R_q) \approx (C_d + C_q) \cdot R_d$ (as opposed to the claim in \cite{5341428,ACERBI201916}), because the fast current flowing through $C_q$ induced by $\mathrm{d}U/\mathrm{d}t$ of $C_d$ is orders of magnitude larger and quicker. This is further proved by figure \ref{fig:model_exp_mismatch}. In \cite{Cova:96}, the authors observed $i_m$ had much sharper leading edge than $i_{out}$ and this was attributed to the fast current flowing through $C_s$.

In our case study of the cumulative signal of SiPM, we use figure \ref{fig:basic_elec_model}, with random dark count, afterpulse and crosstalk as additional non-ideal factors of the waveform. In our case study of the single photon signal of SiPM, we use figure \ref{fig:switching_vol_model} as the SPAD circuit, with all passive circuits ($R_q^{\prime}$, $C_q^{\prime}$, $C_d^{\prime}$) and global capacitance ($C_g$) to form the whole device model (figure \ref{fig:device_multi_model}).

\section{An overview of timing algorithms}
\label{sec:overview_timing_algo}

\begin{figure*}[htb]
	\centering
	\begin{subfigure}[b]{0.3\textwidth}
		\centering
		\includegraphics[width=\linewidth]{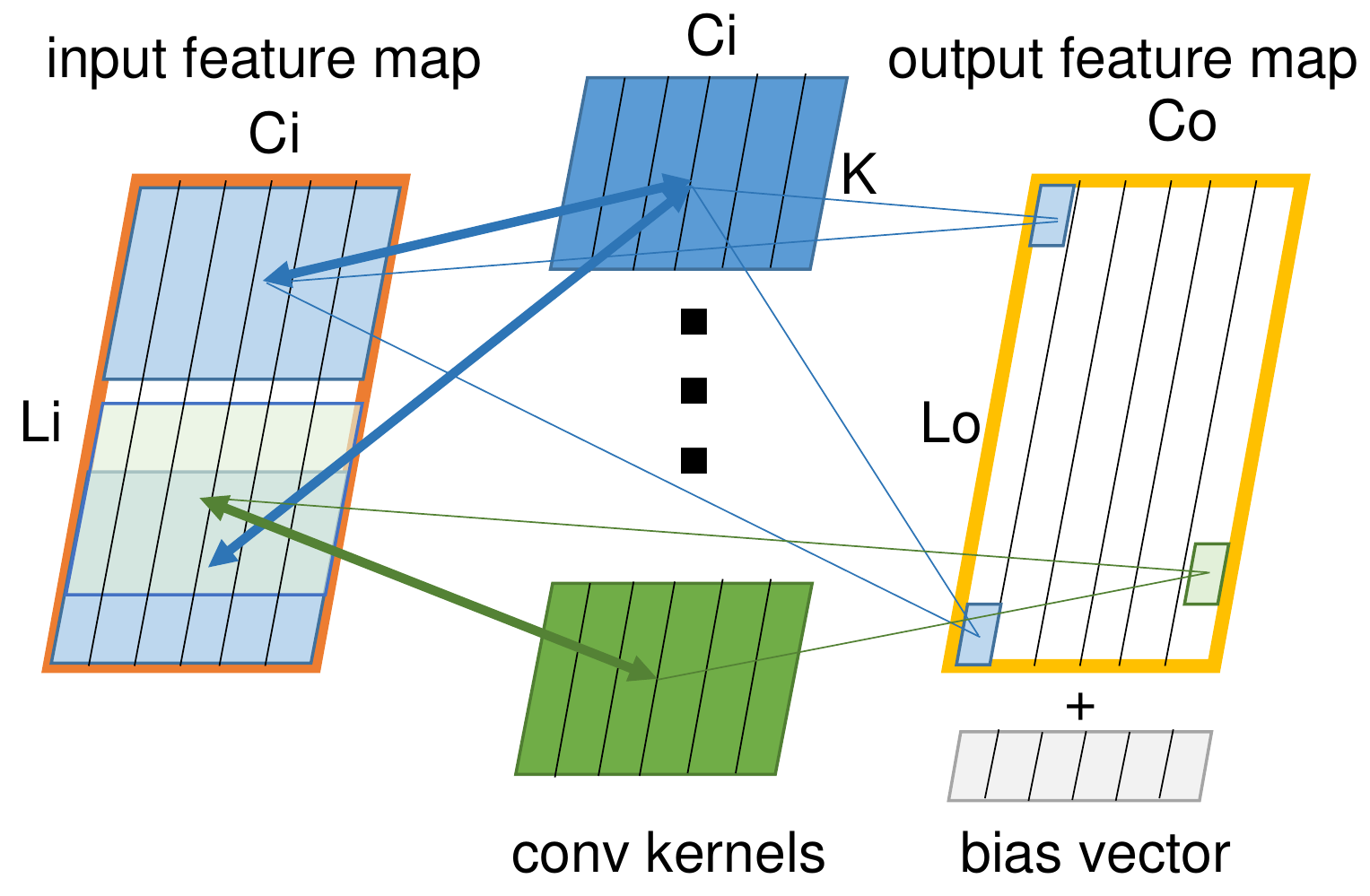}
		\caption{1d convolution}
		\label{fig:conv1d}
	\end{subfigure}
	\begin{subfigure}[b]{0.3\textwidth}
		\centering
		\includegraphics[width=\linewidth]{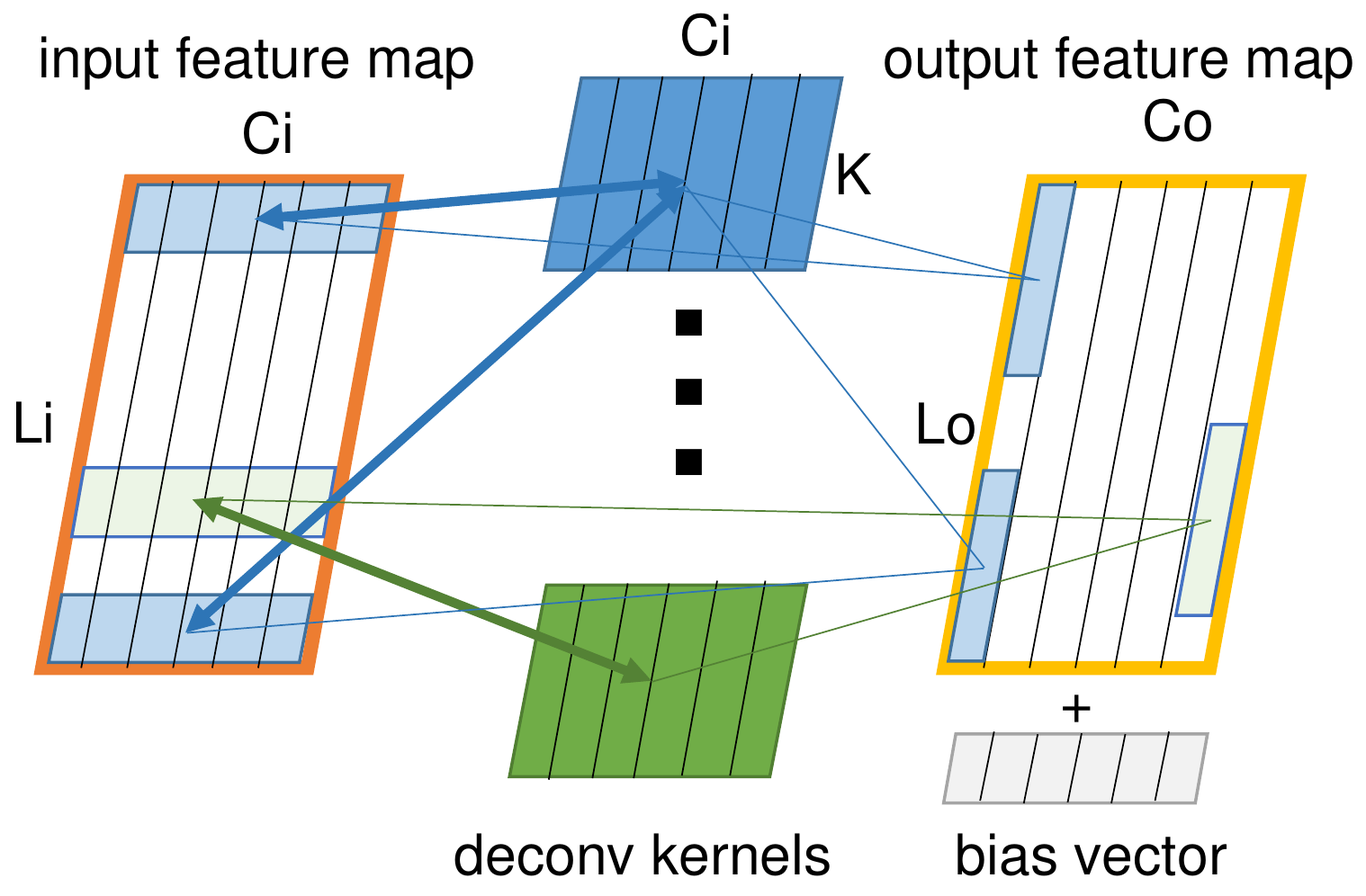}
		\caption{1d deconvolution}
		\label{fig:deconv1d}
	\end{subfigure}
	\begin{subfigure}[b]{0.3\textwidth}
		\centering
		\includegraphics[width=\linewidth]{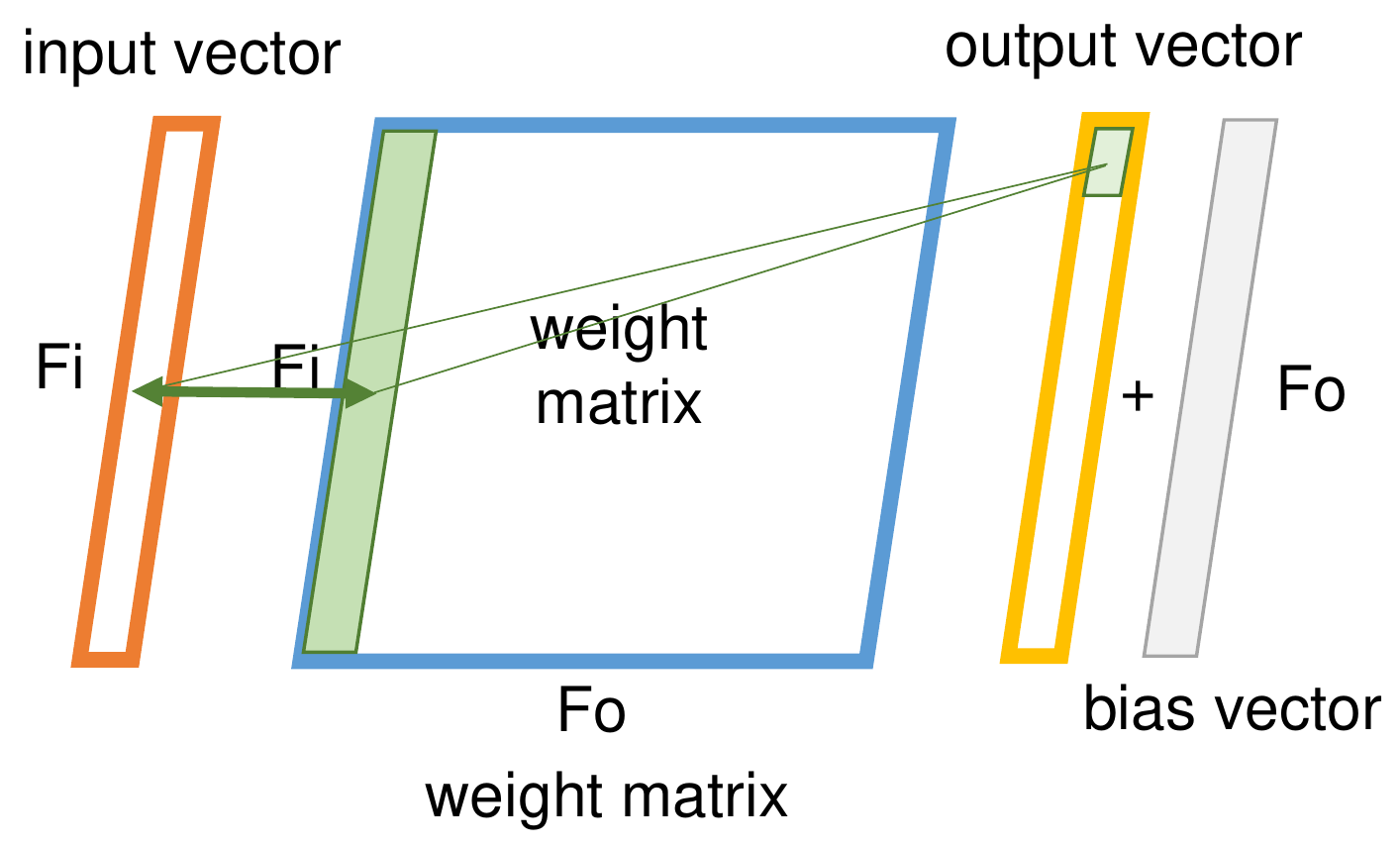}
		\caption{Fully-connected}
		\label{fig:fc1d}
	\end{subfigure}
\caption{Building blocks of one-dimensional CNNs. (a) In 1d (one-dimensional) convolution, each kernel is convolved with a segment of the input feature map across all input channels to generate a single result in an output channel. Per-channel bias is added afterwards. (b) In 1d deconvolution, each value in the input feature map is broadcasted to and multiplied with the corresponding kernel. Then the results in the output channel are summed together. Per-channel bias is added afterwards. (c) In fully-connected multiplication, a vector multiplies with a matrix according to rules of linear algebra and bias is added afterwards.}
\label{fig:blocks_of_nn}
\end{figure*}

Neural networks and variants of CFD are used in the simulation. For neural networks, we center on two architectures constructed by one-dimensional convolution/deconvolution layers and fully-connected layers. The operations of these layers are shown in figure \ref{fig:blocks_of_nn}. CNNs have a moderate amount of parameters and are dominated by localized multiply-and-accumulate operations. These characteristics render them the possibility for efficient hardware implementation. CFD is a well-developed timing method with self-adaptive thresholds to solve the time walk issue of leading edge discrimination. CFD can also be implemented efficiently by digital logic.

\begin{figure*}[htb]
	\centering
	\includegraphics[width=0.85\linewidth]{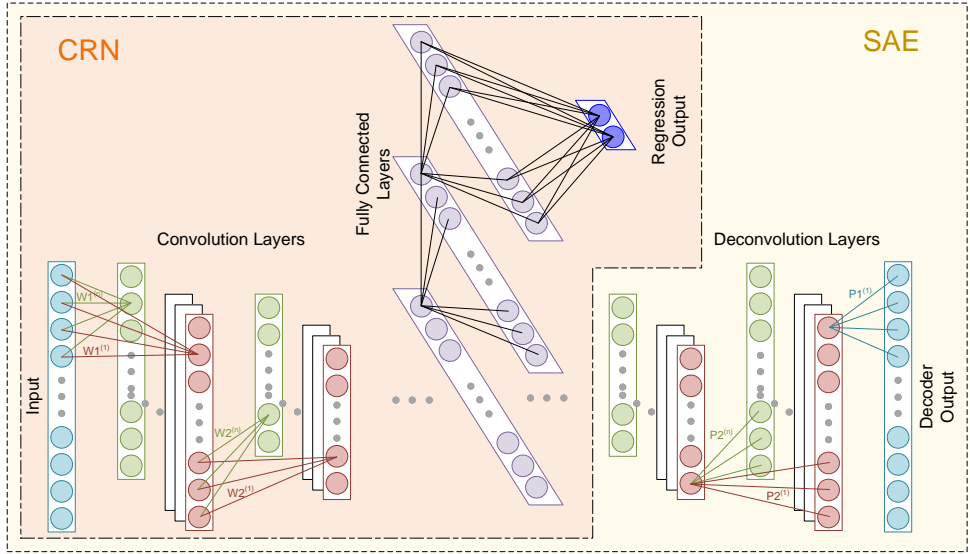}
	\caption{The functional diagram of one-dimensional convolutional regression network and supervised auto-encoder.}
	\label{fig:one_dim_conv}
\end{figure*}

\subsection{Convolutional regression network (CRN)}

The architecture of CRN used in this paper is shown in the light orange shade in figure \ref{fig:one_dim_conv}. It takes the ADC sampling points as the input and predicts the time of arrival (along with energy if needed) of the incident particle. First, convolution layers use localized convolution operations to reduce the length and increase the channels for efficient extraction of waveform features. Then the fully-connected layers are applied upon the results of the last convolution layer to make matrix multiplication for regression of the desired time (and/or energy). In the training phase, a ground-truth label (the value representing the reality, usually used in machine learning) is provided at the regression output and compared to the network prediction. The discrepancy of the label and the prediction is formulated as a loss function which can be used to fine-tune the parameters of the whole network. In the test phase, the test samples are fed forward to generate test results of the performance metrics.

\subsection{Supervised auto-encoder (SAE)}

The architecture of SAE is shown in the light yellow shade in figure \ref{fig:one_dim_conv}. Compared to CRN, it adds a decoder branch to form \emph{multitask learning}: the bottleneck layer (the shared layer in the middle of convolution and deconvolution) is utilized to predict not only the regression targets but also the denoised input. In contrast to the encoder branch made up of convolution layers, the decoder branch is composed of deconvolution layers to increase the length and reduce the channels. There is some research to demonstrate that, in proper conditions, adding a decoder branch as an unsupervised regularizer to the original regression task should not worsen the results \cite{NEURIPS2018_2a38a4a9}. In the training phase, both the regression output and decoder output are compared to labels to generate the loss function. In the test phase, the regression output can be used solely to reduce the computational cost.

\subsection{CFD with zero crossing (CFD-ZC)}

CFD-ZC implements the basic version of the traditional constant fraction discrimination in the digital domain. The input time series of waveform sampling points is divided into two branches. One branch scales the waveform by a diminishing ratio between 0 and 1. The other branch postpones the waveform by a fixed delay. Then the latter branch is taken the negative of and added to the former branch. The zero-crossing point of the differential waveform is regarded as the timing result. Usually this point does not coincide with any sampling points. Then a linear interpolation is performed between the nearby two sampling points to compute the adjusting value with respect to the sampling time period.

\subsection{CFD with maximum amplitude (CFD-MAX)}

CFD-MAX is the widely used version in the literature and usually termed as digital constant fraction discrimination (dCFD). It works by finding the maximum value of the input time series of sampling points and choosing a threshold with a fixed diminishing ratio (eg. 0.5) relative to the maximum value. The timing result is computed on the original waveform regarding the threshold and interpolated between the nearby two sampling points. Compared to CFD-ZC, CFD-MAX is more friendly to digital logic, and can be implemented with digital integrated circuits.

\subsection{CFD with interpolation (CFD-INT)}

In recent works \cite{8081803,9187849}, real-time interpolation is proposed to be used before dCFD to improve the estimation of both the waveform maximum value and threshold crossing point. This variant is termed as CFD-INT in this paper. The interpolating process is in essence zero-stuffing the sampling points and lowpass filtering. After interpolation, the equivalent sampling frequency is several times larger than the original waveform so that the effect of randomness in the sampling process is reduced. It should be noted that interpolation does not provide additional information of the sampling points and only works as an auxiliary measure to improve the performance of dCFD.

\section{Simulation results}

In this section, two application scenarios are studied with both the computed lower bound and timing algorithms. Then a discussion about the implication of simulation results is presented. The simulation time step is set to 10 ps. The intensity of noise is represented by the standard deviation ($\sigma$) of each individual sampling point. For a full explanation of the noise term in the mathematical model, refer to first two sections in the appendix. For each pixel (with particular critical frequency and noise level) in the two-dimensional image, we train neural networks from scratch and test them for the resolution criterion. The configurations for training and testing are described in section \ref{sec:NN_conf}, and the network architecture is given in section \ref{sec:NN_arch}. For 800 MHz sampling rate, we use the 64-point networks (table \ref{tab:hyper_64_points}); for 200 MHz sampling rate, we use the 16-point networks (table \ref{tab:hyper_16_points}).

\begin{figure*}[htb]
	\centering
	\begin{subfigure}[b]{0.3\textwidth}
		\centering
		\includegraphics[width=\linewidth]{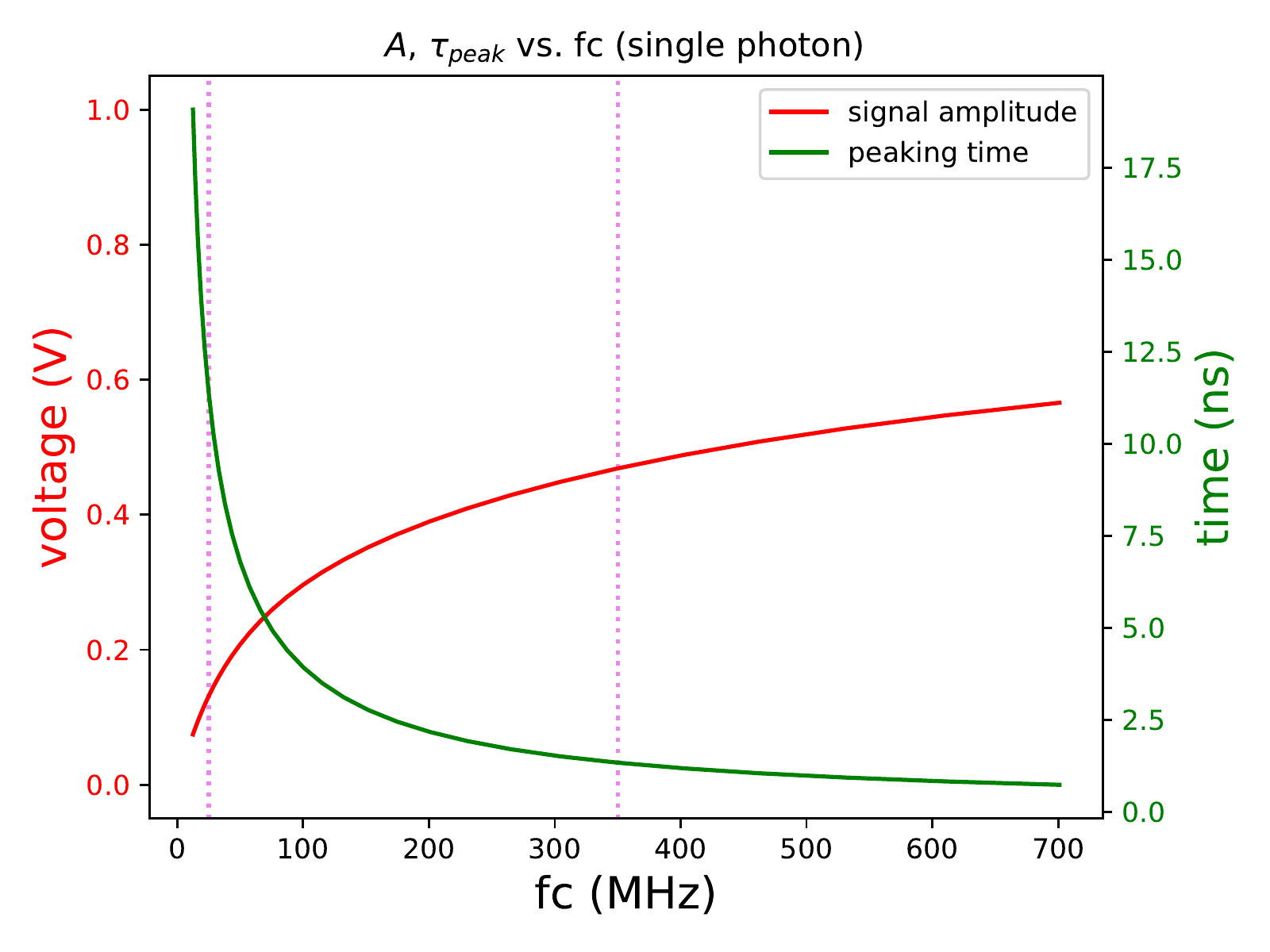}
		\caption{Single photon detection}
		\label{fig:amp_peaking_sp}
	\end{subfigure}
	\begin{subfigure}[b]{0.3\textwidth}
		\centering
		\includegraphics[width=\linewidth]{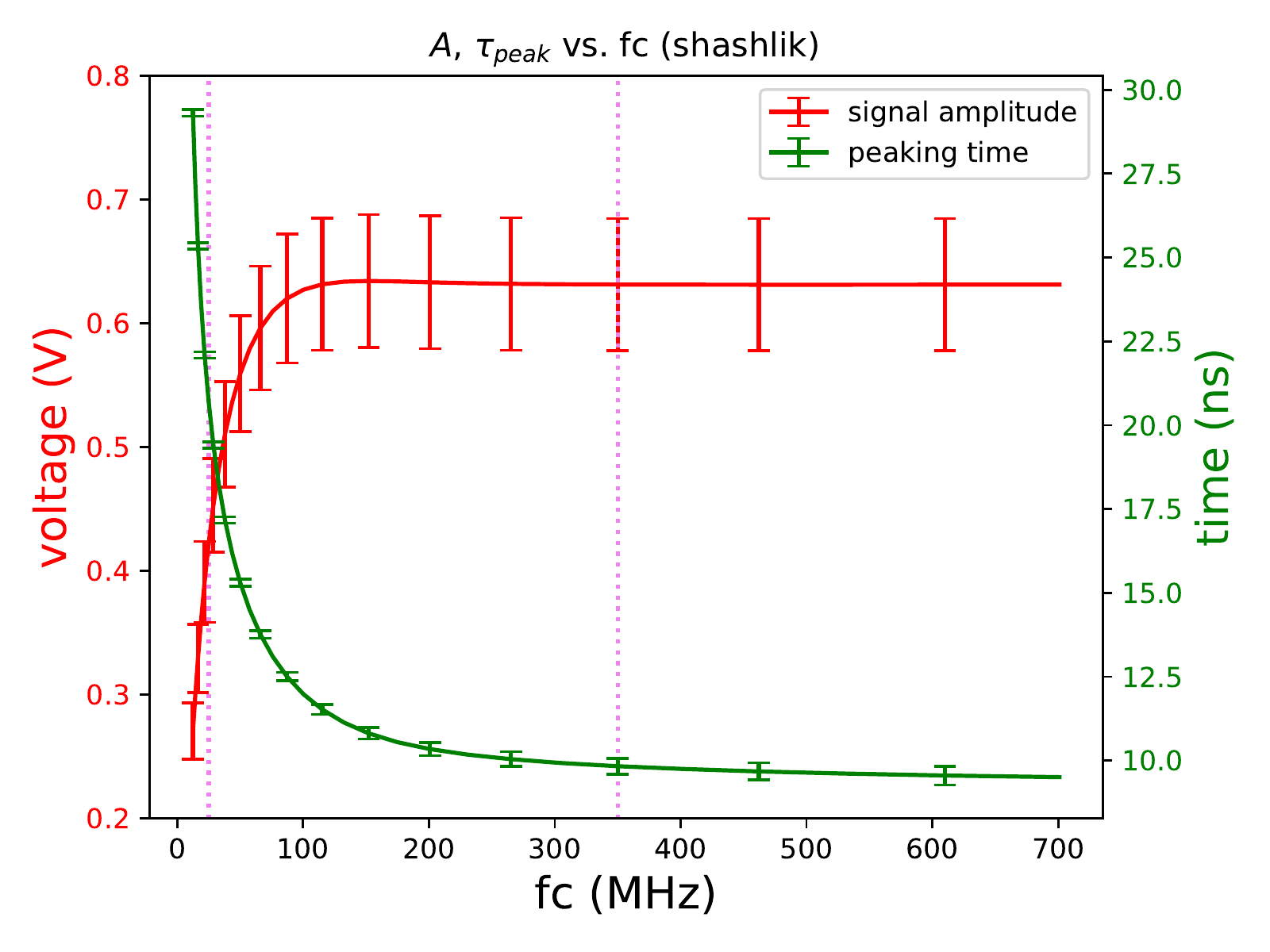}
		\caption{Shashlik-type calorimeter}
		\label{fig:amp_peaking_shashlik}
	\end{subfigure}
	\begin{subfigure}[b]{0.3\textwidth}
		\centering
		\includegraphics[width=\linewidth]{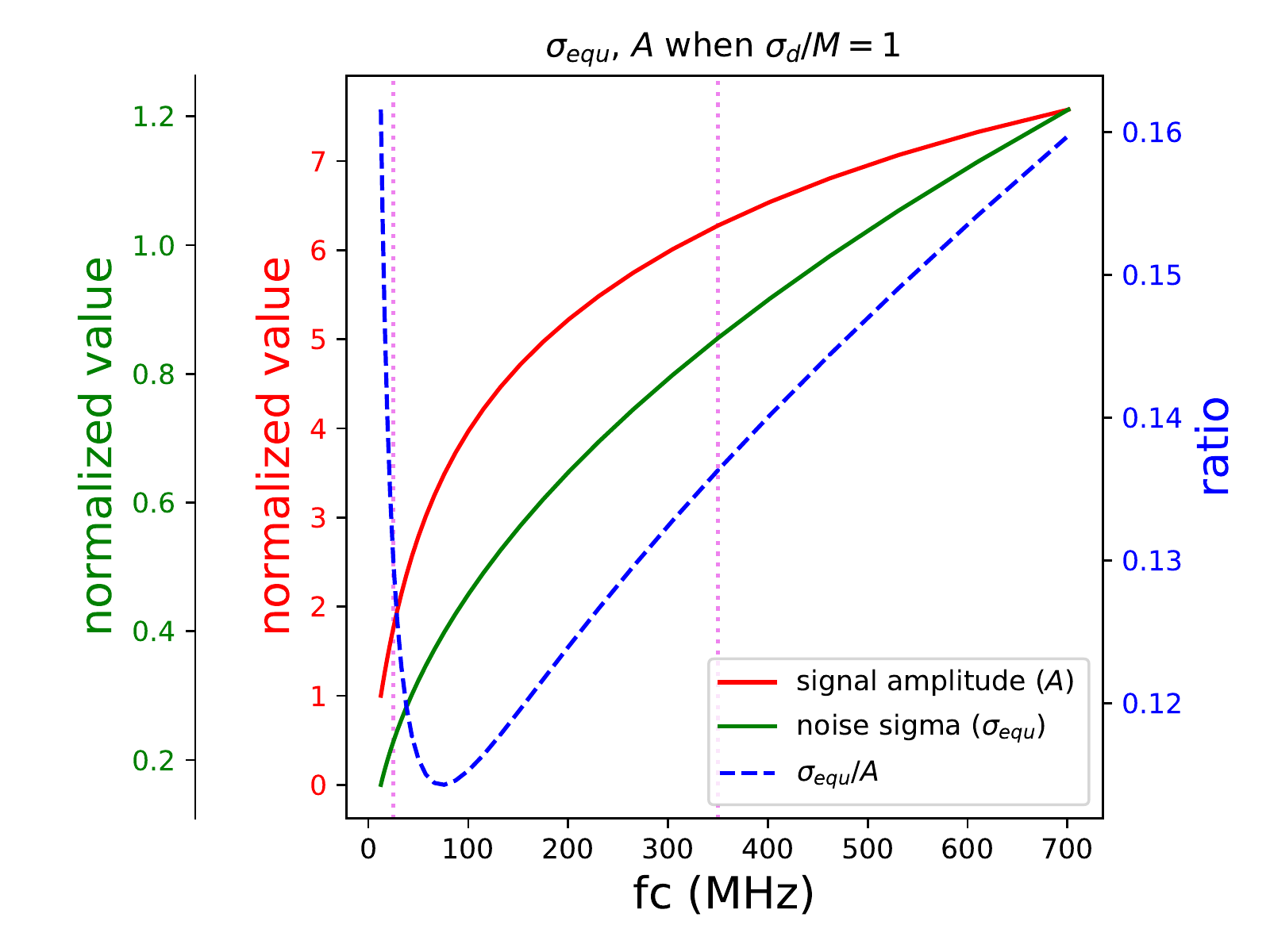}
		\caption{Single photon detection}
		\label{fig:equiv_noise_at_term}
	\end{subfigure}
	\caption{Signal characteristics affected by the analog channel. The purple dotted lines indicate the researched range of critical frequency. (a) Amplitude and peaking time for single photon detection when variables are fixed at most probable values. (b) Amplitude and peaking time for shashlik-type calorimeter. The mean and standard deviation (error bars) are statistical results on the simulation dataset. (c) The equivalent noise level at the waveform sampling side caused by the source side. The correlation of different sampling points is not plotted in the figure.}
	\label{fig:single_photon_prepare}
\end{figure*}

\subsection{Application 1: single photon detection}
\label{sec:app_1}

\begin{table}[htb]
	\centering
	\begin{tabular}{ccccccc} 
		\hline
		& $R_d$ & $C_d$ & $V_{br}$ & $R_q$ & $C_q$ & $C_g$ \\
		\hline
		mean & 1.0 \si{\kilo\ohm} & 15.0 \si{fF} & 68.8 \si{V} & 179 \si{\kilo\ohm} & 4.30 \si{fF} & 7.50 \si{pF} \\ 
		std. & -- & 0.612 \si{fF} & 0.765 \si{V} & 510 \si{\ohm} & 0.612 \si{fF} & 0.969 \si{pF} \\
		\hline
	\end{tabular}
	\caption{Mean and standard deviation of variables used in the simulation of single photon waveform. These values are converted from the model parameters with 95\% confidence intervals in \cite{5341428} (SiPM: MPPC-S10362-11-25u). The bias voltage is fixed at 71.1 \si{V}.}
	\label{tab:single_photon_variables}
\end{table}

The simulation starts by sampling parameters from a multivariate Gaussian distribution. To account for variation of parameters, we consult experimental values in \cite{5341428} and use them as intrinsic variables, shown in table \ref{tab:single_photon_variables}. Each variable is independent, so the covariance matrix is a diagonal matrix with diagonal elements being variance of variables. Only the fired (active) cell in SiPM is randomly sampled; the passive circuits of unfired cells do not vary because their collective variance is negligible.

The closed form of the waveform function is computed by Laplace transformation, and the system of linear equations is solved in the s-domain. The quenching behavior is simulated by a switch opened at a fixed quenching current (30 \si{\micro\ampere}). The analog channel is abstracted as a second-order lowpass filter with a variable critical frequency. Figure \ref{fig:amp_peaking_sp} shows the amplitude and peaking time of the waveform after the analog channel. The researched range of critical frequency is from 25 MHz to 350 MHz on logarithmic scale. It can be seen that the critical frequency affects the slope and time period of the leading edge. In the considered range, the duration of leading edge is less than 12.5 \si{ns} and corresponds to less than 10 sampling points at the 800 MHz sampling rate.

If we denote the most probable amplitude of the original signal as $M$, the noise level $\sigma_d / M$ at the source side is chosen in the range between 0.01 and 1.4 on logarithmic scale. After lowpass filtering, the relative noise level will diminish. Figure \ref{fig:equiv_noise_at_term} shows the equivalent noise level $\sigma_{equ} / A$ at the waveform sampling side (where $A$ is the most probable amplitude) caused by the source side when $\sigma_d / M$ is fixed at 1. The corresponding range of $\sigma_{equ} / A$ is approximately from 0.001 to 0.2 when $\sigma_d / M$ varies. It should be noted that the correlation of different sampling points is not represented by the figure, and the actual performance will drop because of correlation. Besides, the additional noise $\sigma_s / A$ at the terminal is fixed at 0.001.

\begin{figure*}[htb]
	\centering
	\includegraphics[width=0.95\linewidth]{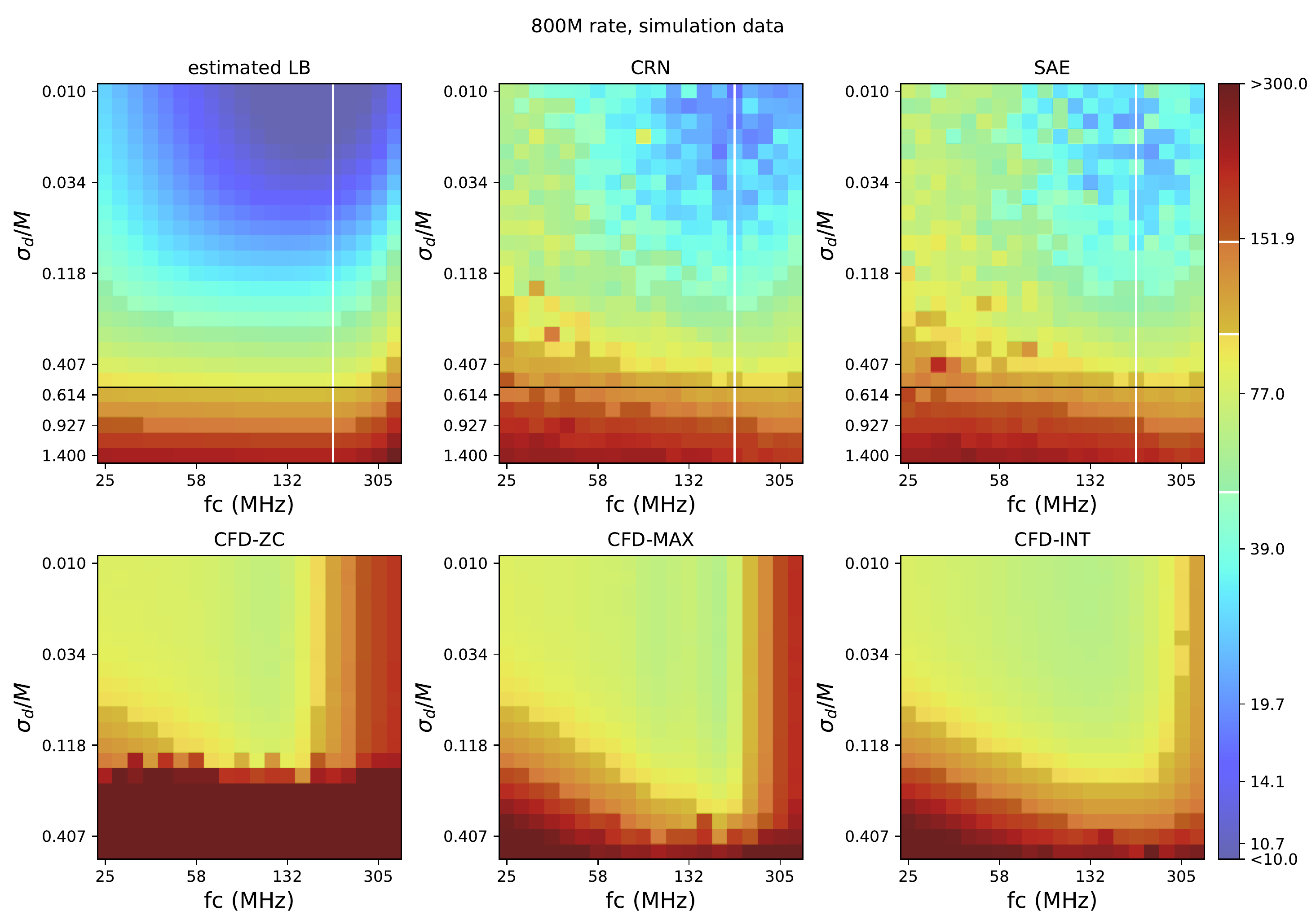}
	\caption{The estimated lower bound and results from different timing algorithms in the 5-variable condition of single photon waveform. In each sub-figure, the timing resolution (in unit \si{ps}) is plotted versus different noise levels at the source side and critical frequency of the analog channel. For CFDs, we omit the conditions of extremely high noise level because they do not exhibit advantage in this region. Four different shades indicate four regions: $<$50 ps, 50 ps$\sim$100 ps, 100 ps$\sim$150 ps and $\geq$150 ps. The white lines on the images indicate the critical frequency used in the line plot below.}
	\label{fig:single_photon_default_compare_all}
\end{figure*}

In figure \ref{fig:single_photon_default_compare_all}, we show the estimated lower bound and results from different timing algorithms in the 5-variable condition. The computation of lower bound is conducted with the formulation and parametrization stated in section \ref{sec:sp_signal_sipm}. In this figure and figures below, we use logarithmic coordinates for noise level, critical frequency and timing resolution. In the top left sub-figure, it can be seen that the estimated achievable performance changes monotonously with the noise level but not with the critical frequency. Low critical frequency tends to decrease the slope of the leading edge and high critical frequency makes the sampling points few and inconstant.

The top center and top right sub-figures give the results from neural networks. Some fluctuation resides in the resolution due to the random training process of neural networks. For CRN and SAE, there is a region at high critical frequency and low noise level where the timing resolution is better than 50 ps. The three bottom sub-figures give the results from traditional timing methods. CFDs are unable to achieve the resolution below 50 ps and show serious deterioration in performance at high noise level. Relatively, neural networks are suited to the high critical frequency region where the sampling of leading edge is insufficient, and CFDs are suited to the low critical frequency region where the slope of leading edge is flattened. Besides, the overall tendency of the lower bound at the researched range of critical frequency and noise level is more consistent with neural networks than CFDs, except for some mismatches at high critical frequency and extremely high noise level.

\begin{figure*}[htb]
	\centering
	\includegraphics[width=0.95\linewidth]{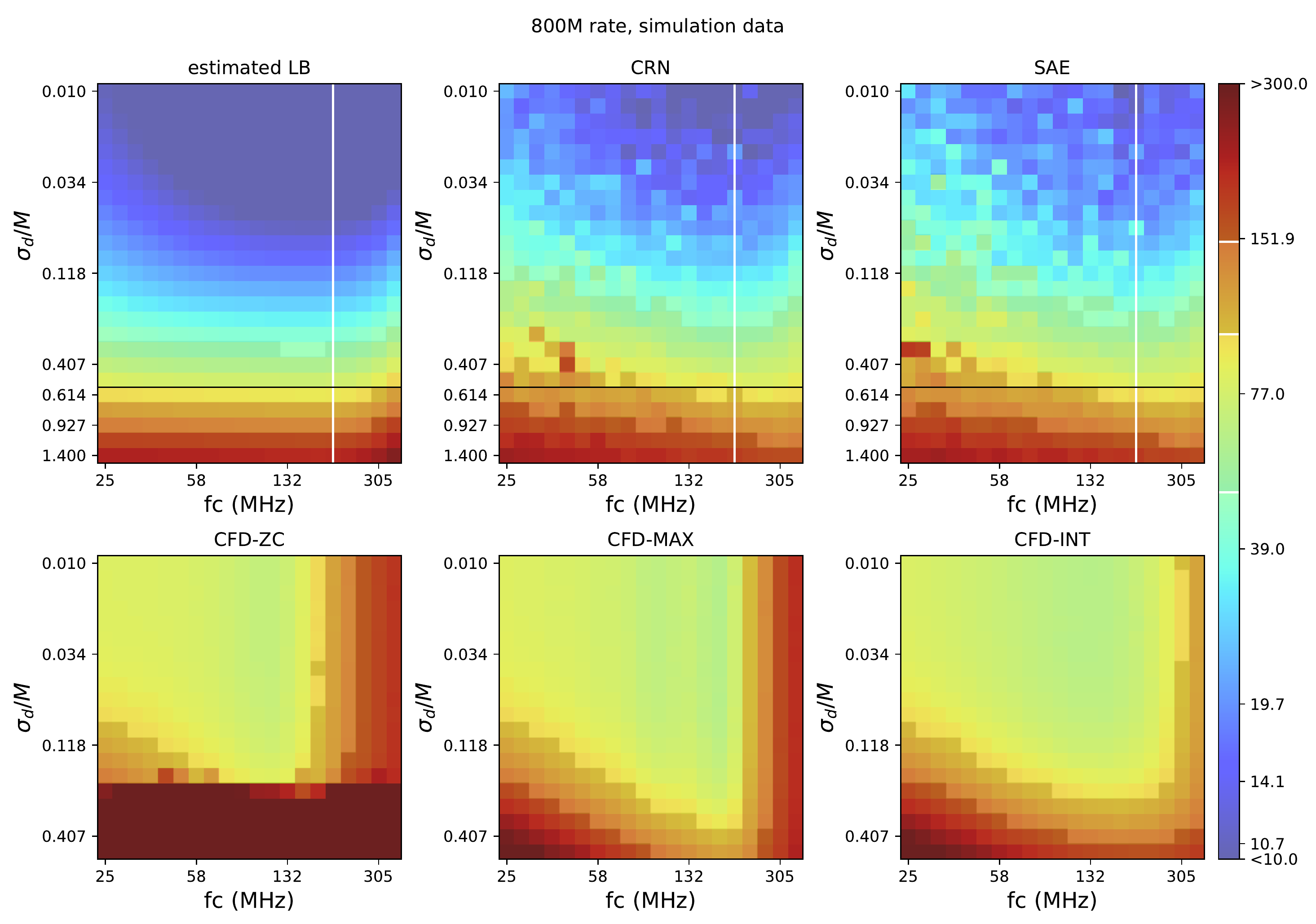}
	\caption{The estimated lower bound and results from different timing algorithms in the 4-variable condition (without $V_{br}$) of single photon waveform. In each sub-figure, the timing resolution (in unit \si{ps}) is plotted versus different noise levels at the source side and critical frequency of the analog channel. For CFDs, we omit the conditions of extremely high noise level because they do not exhibit advantage in this region. Four different shades indicate four regions: $<$50 ps, 50 ps$\sim$100 ps, 100 ps$\sim$150 ps and $\geq$150 ps. The white lines on the images indicate the critical frequency used in the line plot below.}
	\label{fig:single_photon_wo_vbr_compare_all}
\end{figure*}

In table \ref{tab:single_photon_variables}, the standard deviation of $V_{br}$ is very significant considering the over-voltage (71.1 - 68.8 = 2.3 \si{V}). This is too pessimistic for modern SiPM devices whose variation of the gain is almost unity. Thus, we further study the 4-variable condition when $V_{br}$ is fixed to its mean value. The result is shown in figure \ref{fig:single_photon_wo_vbr_compare_all}. In the top left sub-figure, we can see a significant improvement of the computed lower bound reaching sub-10 \si{ps} performance at low noise level. Neural networks, meanwhile, improve with the lower bound and have a very similar distribution. In contrast, the improvement of CFDs is very limited to the region with high noise level and does not probe into the region where they have the best performance. The deviation of CFDs shows their limitations essentially being regardless of the prior distribution of the signal waveform. Neural networks, on the other hand, can always effectively make use of the data and incorporate the information from the global distribution of signal waveform into its model.

\begin{figure*}[htb]
	\centering
	\begin{subfigure}[b]{0.45\textwidth}
		\centering
		\includegraphics[width=\linewidth]{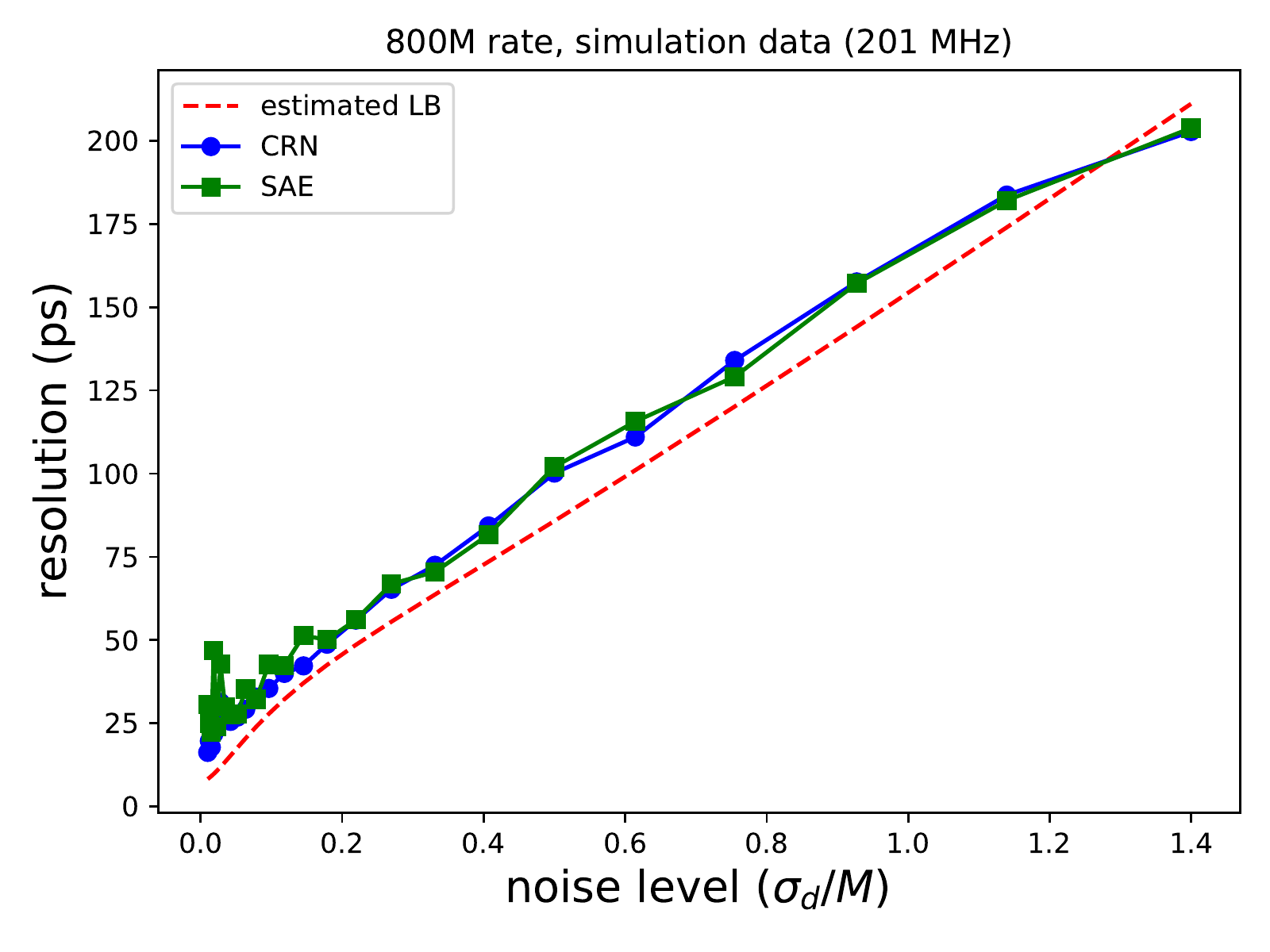}
		\caption{5-variable condition}
		\label{fig:sp_default_line}
	\end{subfigure}
	\begin{subfigure}[b]{0.45\textwidth}
		\centering
		\includegraphics[width=\linewidth]{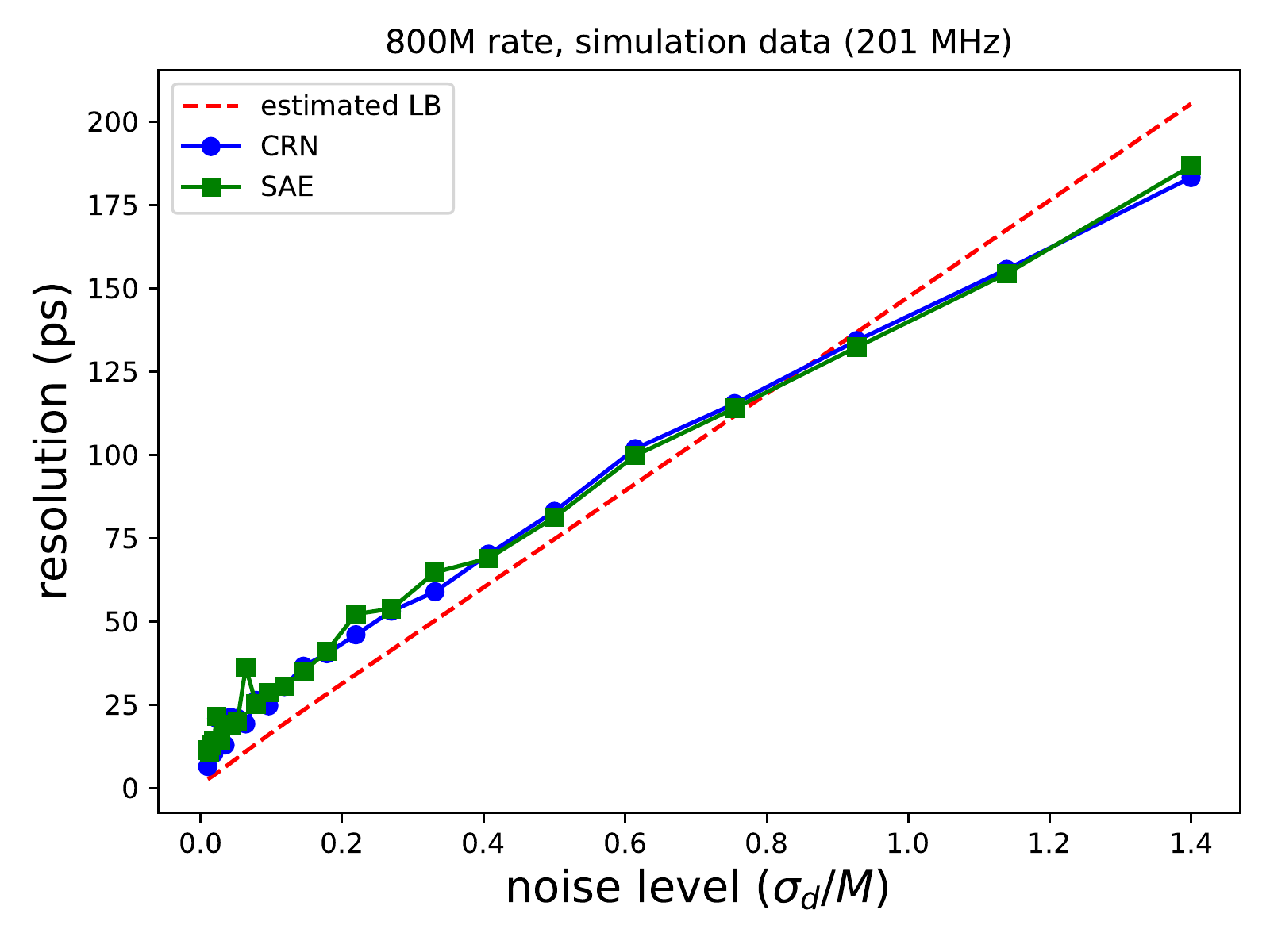}
		\caption{4-variable condition}
		\label{fig:sp_wo_vbr_line}
	\end{subfigure}
	\caption{Comparisons between the estimated lower bound and neural networks at fixed critical frequency (201 MHz) in the (a) 5-variable condition and (b) 4-variable condition.}
	\label{fig:sp_line_plot}
\end{figure*}

In figure \ref{fig:sp_line_plot}, we extract the results from images to compare the timing performance of neural networks with the lower bound. Generally the overall trend of neural networks matches the estimation of lower bound quite well. We notice that at high noise levels the lower bound tends to overestimate the achievable resolution. This is reasonable since we only compute the lower bound at most probable values of intrinsic parameters (see the last paragraph in section \ref{sec:model_compute_lb}). A more accurate lower bound can be acquired by Monte Carlo simulation to marginalize all parameters. We believe the estimated lower bound is informative and directive considering the limited computing power.

\subsection{Application 2: shashlik-type calorimeter}
\label{sec:app_2}

\begin{table}[htb]
	\centering
	\begin{tabular}{cccccccc} 
		\hline
		& $R_d$ & $C_d$ & $V_{br}$ & $R_q$ & $F_d$ & $P_{ap}$ & $P_c$ \\
		\hline
		value & 1.0 \si{\kilo\ohm} & 22.4 \si{fF} & 53.0 \si{V} & 300 \si{\kilo\ohm} & 1600 kcps & 3 \% & 1 \% \\
		\hline
	\end{tabular}
	\caption{Values of variables used in the simulation of shashlik-type calorimeter. These values are acquired from the official datasheet (SiPM: MPPC-S13360-6025PE). $F_d$, $P_{ap}$ and $P_c$ represent the dark count rate (in unit kilo counts per second), afterpulse probability and crosstalk probability respectively. The bias voltage is fixed at 58.0 \si{V}.}
	\label{tab:shashlik_variables}
\end{table}

In each event, we record the arrival time and position of each photon being effectively detected. The SiPM signal pulse is generated with the parameters listed in table \ref{tab:shashlik_variables}. The researched range of critical frequency is from 25 MHz to 350 MHz, and noise level $\sigma_w / A$ at the waveform sampling side (where $A$ is the most probable amplitude) is from 0.001 to 0.05, both on logarithmic scale.

\begin{figure*}[htbp]
	\centering
	\includegraphics[width=0.95\linewidth]{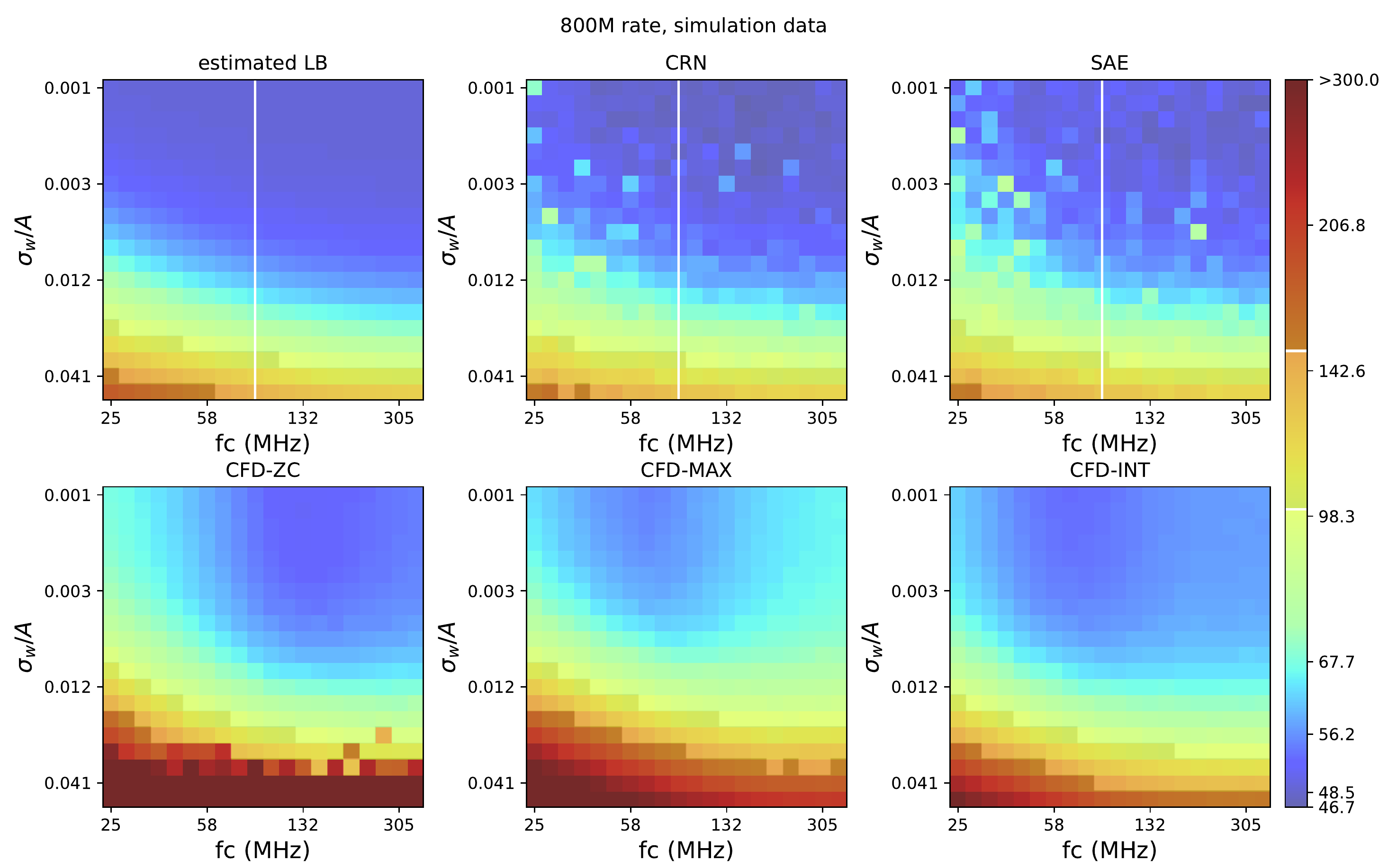}
	\caption{The estimated lower bound and results from different timing algorithms for the shashlik-type calorimeter at the 800 MHz sampling rate. In each sub-figure, the timing resolution (in unit \si{ps}) is plotted versus different noise levels and critical frequency of the analog channel. The color palette is adjusted to emphasize the regions with small values. Three different shades indicate three regions: $<$100 ps, 100 ps$\sim$150 ps and $\geq$150 ps. The white lines on the images indicate the critical frequency used in the line plot below.}
	\label{fig:ab_base_compare_all}
\end{figure*}

\begin{figure}[htbp]
	\centering
	\includegraphics[width=0.45\linewidth]{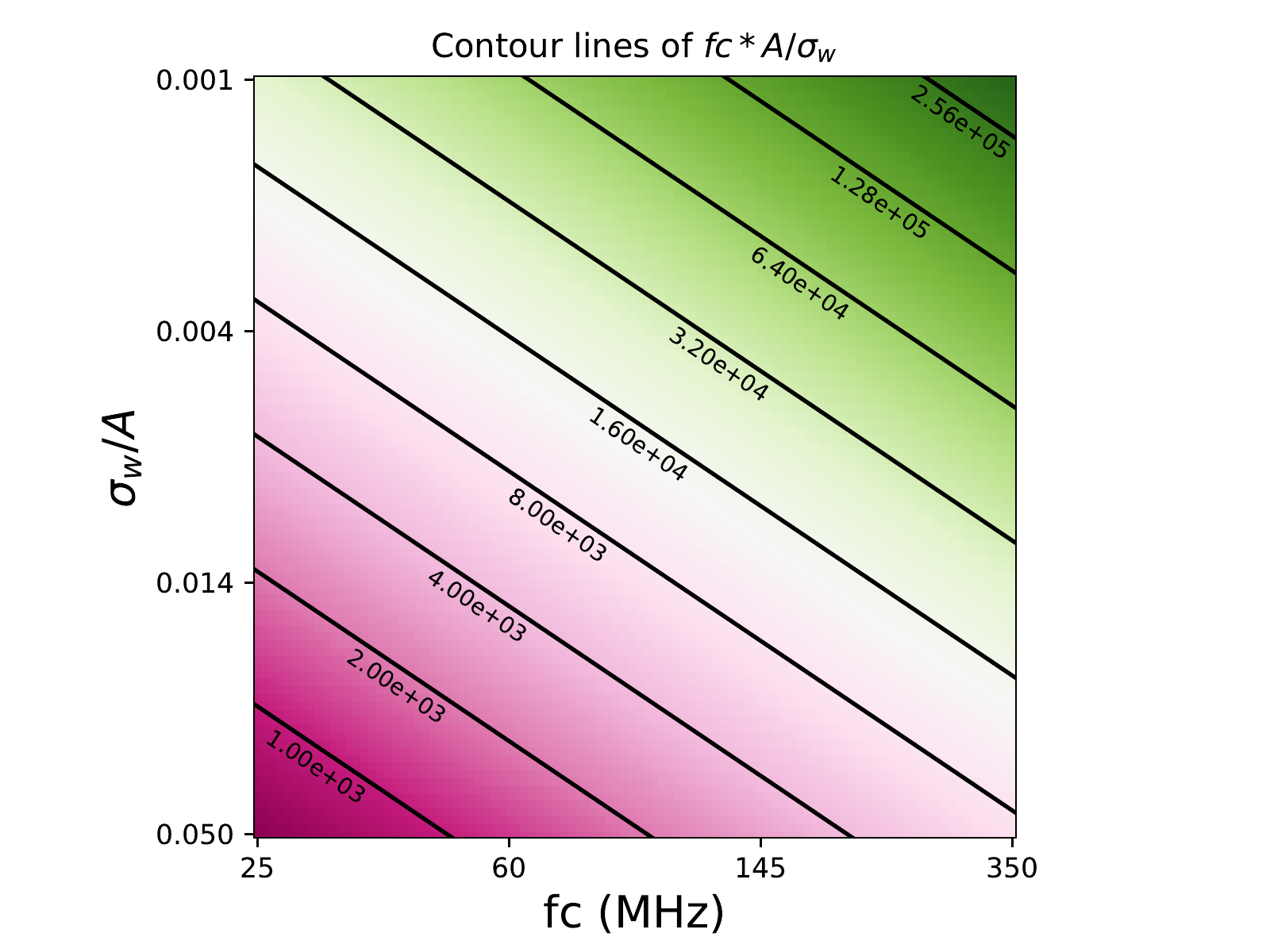}
	\caption{The spatial distribution of $fc * A / \sigma_w$ in the researched range of noise level and critical frequency. The values are plotted on logarithmic scale, with contour lines indicating specific thresholds.}
	\label{fig:ratio_blk}
\end{figure}

In figure \ref{fig:ab_base_compare_all}, we show the estimated lower bound and results from different timing algorithms when sampling the waveform at the 800 MHz sampling rate. We follow the modelling method stated in section \ref{sec:cum_signal_sipm} to compute the lower bound. Due to the intrinsic rise time of detector signal (usually in the order of several \si{ns}), sampling at 800 MHz is high enough to keep the details of the leading edge after passing through the analog channel. This fact is shown in figure \ref{fig:amp_peaking_shashlik} where the peaking time is well above 10 \si{ns} in the researched range. Thus in the top left sub-figure, the best achievable performance changes monotonously with the noise level and also with the critical frequency.

In the top center (CRN) and top right (SAE) sub-figures, the results from neural networks have very similar trends with the lower bound in spite of some fluctuations. There is a considerably large region at low noise level where the lower bound and neural networks do not have apparent difference in performance (50 \si{ps} or less). In the three bottom sub-figures, the results from CFDs are not as good as neural networks but still comparable. CFD-ZC achieves better resolution at relatively high critical frequency, while CFD-MAX and CFD-INT are more competent at critical frequency around 66 MHz. For noise level below 0.01, all the timing algorithms attain the resolution below 100 \si{ps} at almost any critical frequency.

\begin{figure*}[htb]
	\centering
	\includegraphics[width=0.95\linewidth]{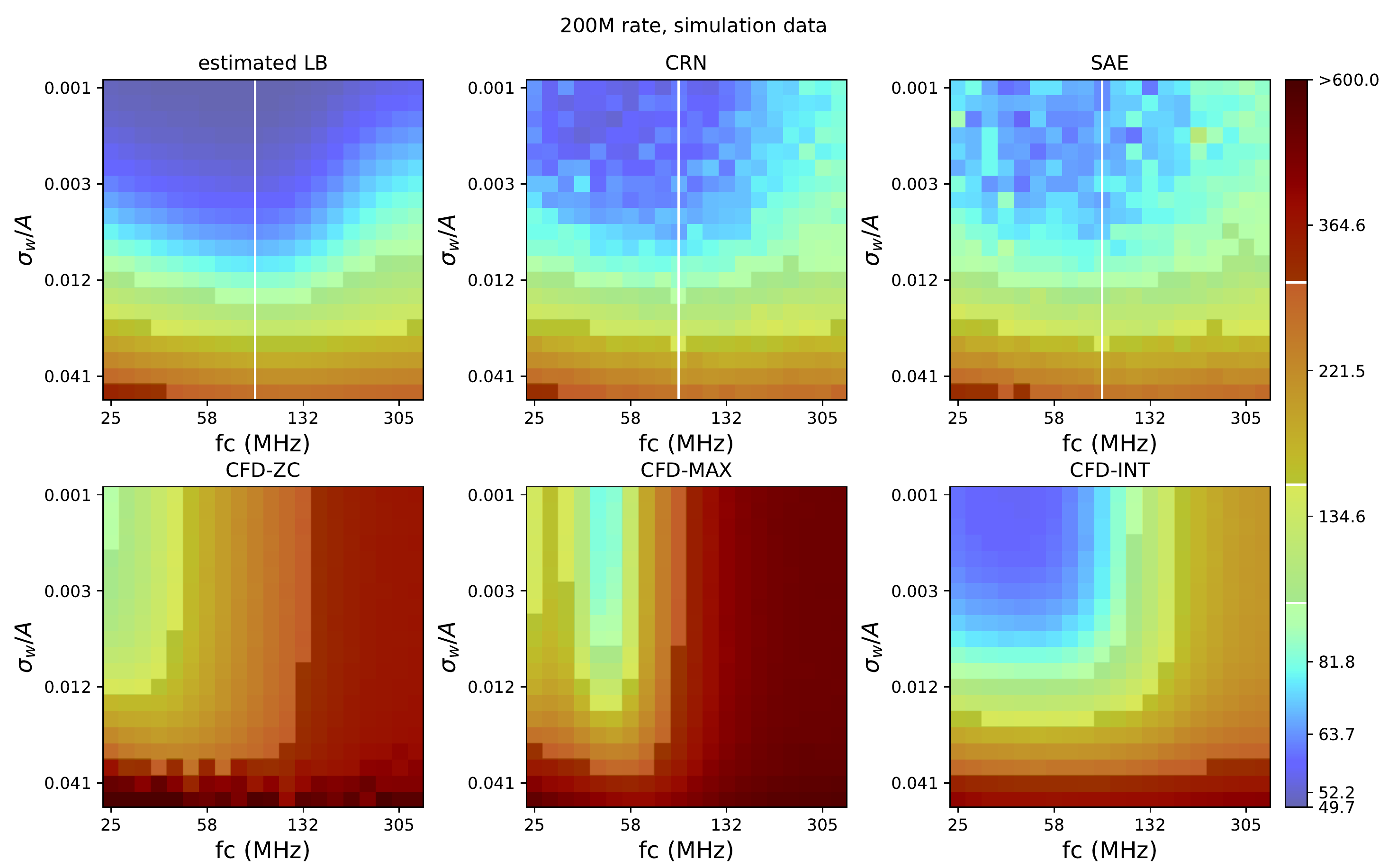}
	\caption{The estimated lower bound and results from different timing algorithms for the shashlik-type calorimeter at the 200 MHz sampling rate. In each sub-figure, the timing resolution (in unit \si{ps}) is plotted versus different noise levels and critical frequency of the analog channel. The color palette is adjusted to emphasize the regions with small values. Four different shades indicate four regions: $<$100 ps, 100 ps$\sim$150 ps, 150 ps$\sim$300 ps and $\geq$300 ps. The white lines on the images indicate the critical frequency used in the line plot below.}
	\label{fig:ab_16p_base_compare_all}
\end{figure*}

In engineering applications, such as high energy physics experiments, sometimes it is impractical to use ADCs with 800 MHz or higher sampling rates. To investigate the behavior and limitation of timing algorithms at low sampling rates, we further study the case with the 200 MHz sampling rate, shown in figure \ref{fig:ab_16p_base_compare_all}. In the top left sub-figure, we can see the computed lower bound drops at high critical frequency because of insufficient sampling at low rates. However, the best achievable performance can still be around 50 \si{ps} at low noise level.

For CRN and SAE, we observe similar resolution with the lower bound at relatively low critical frequency. At high critical frequency, the best results of neural networks have a small gap but are still comparable. In contrast, CFDs are more susceptible to the change of sampling rates. For CFD-ZC and CFD-MAX, only a small region in the researched range reaches the resolution better than 150 \si{ps}, and the deviation from the lower bound is apparent. Nevertheless, CFD-INT is the best of CFD variants and show competitive results at low critical frequency. Finally, it should be noted that the lower bound is also directive for CFD-INT.

\begin{figure*}[htb]
	\centering
	\begin{subfigure}[b]{0.45\textwidth}
		\centering
		\includegraphics[width=\linewidth]{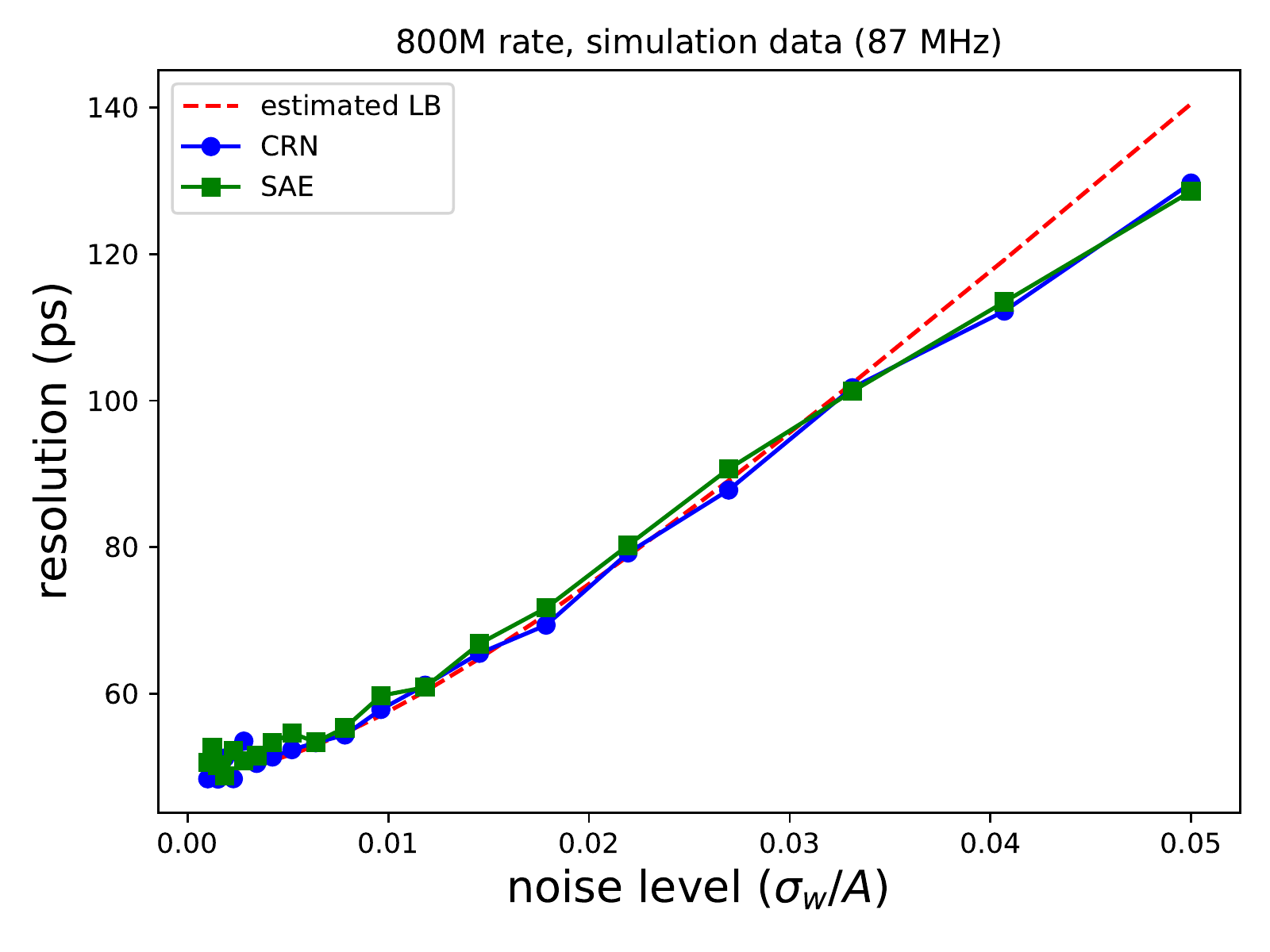}
		\caption{800 MHz sampling rate}
		\label{fig:ab_base_line}
	\end{subfigure}
	\begin{subfigure}[b]{0.45\textwidth}
		\centering
		\includegraphics[width=\linewidth]{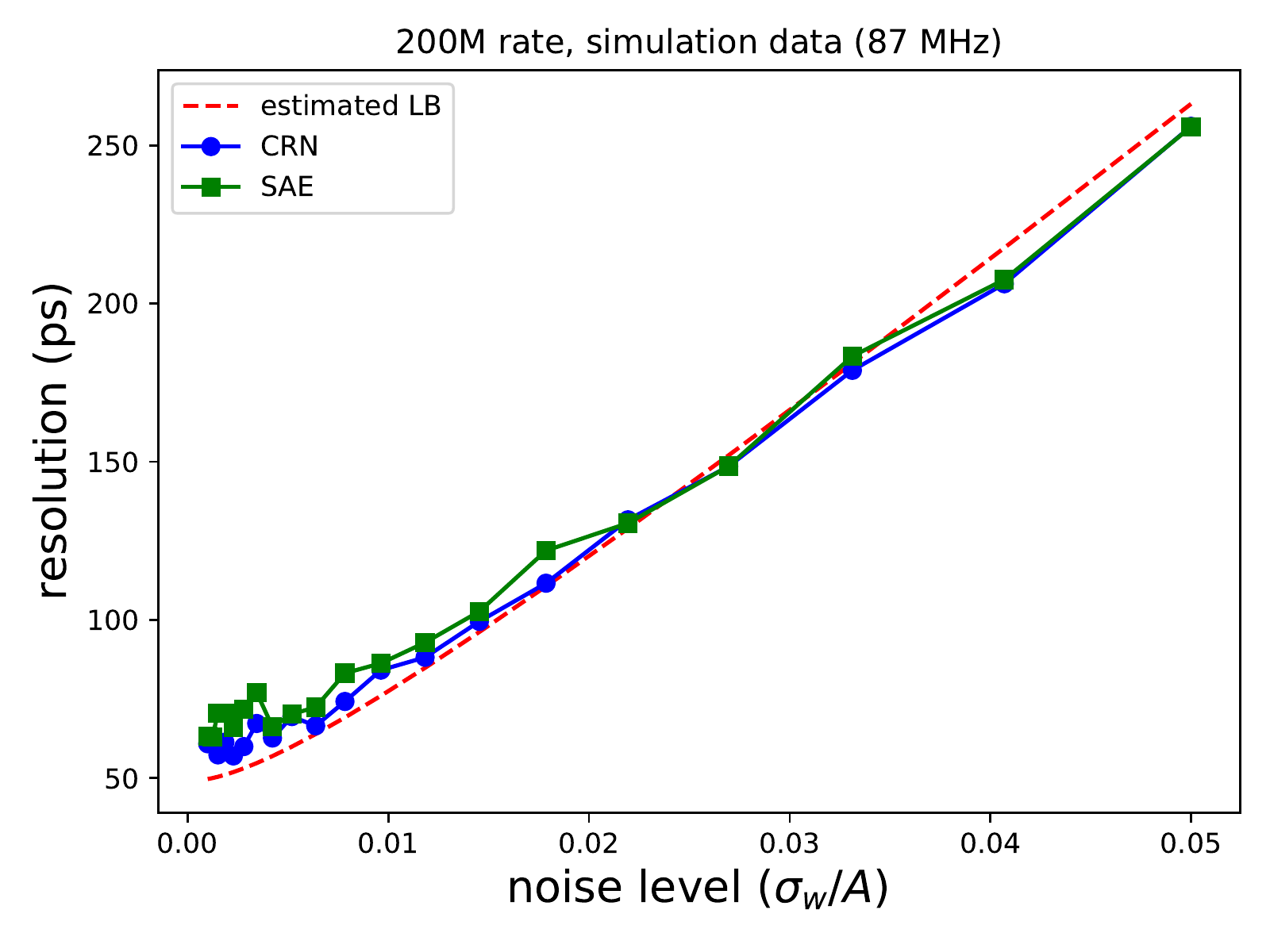}
		\caption{200 MHz sampling rate}
		\label{fig:ab_16p_base_line}
	\end{subfigure}
	\caption{Comparisons between the estimated lower bound and neural networks at fixed critical frequency (87 MHz) at the (a) 800 MHz sampling rate and (b) 200 MHz sampling rate.}
	\label{fig:ab_line_plot}
\end{figure*}

To carefully examine the difference between the lower bound and neural networks, again we plot the results from images at fixed critical frequency, shown in figure \ref{fig:ab_line_plot}. The extents of y-axis in these two subplots demonstrate the advantage of high sampling rates for timing performance. The trend of neural networks is consistent with the estimated lower bound both at the 800 MHz rate and the 200 MHz rate. The overestimation of lower bound at high noise levels is due to the same reason as the single photon case.

\subsection{Discussion}

\paragraph{Parameter optimization} Based on the simulation results, it is meaningful to optimize the parameters to achieve the best performance with reasonable cost. For example, noise is detrimental to the timing resolution in all cases. However, for a certain critical frequency, there is a threshold for the noise level. Reducing noise below the threshold will not significantly improve performance since the spread of waveform variables dominates the resolution in this region. So one should judge the cost-effectiveness when doing low-noise front end circuit design.

Besides, different applications lead to different optimization targets. In scenarios when the counting rate is high and we want better discrimination, we should control the trailing edge of the pulse waveform, so the critical frequency of the analog channel should be set higher when the performance requirements are satisfied. In other scenarios (such as calorimeters) when the counting rate is not the limiting factor, the critical frequency is optimized with a different strategy. In figure \ref{fig:ratio_blk}, the numbers above the black lines indicate the equivalent gain bandwidth product to noise ratio. When the ratio is high, it implies the noise is not prominent in the circuit, so it is feasible to use lower critical frequency (the left extent of black lines). When the ratio is low, it implies the noise is prominent, and there is usually an optimal critical frequency and an optimal algorithm to achieve the best performance.

\paragraph{Algorithm assessment} The above results are also helpful to assess the ability and potential of timing algorithms. In total, CFDs are not consistent with the lower bound mainly because of their working principle. It is meaningful to think of better variants of CFD to exploit more information from data (eg. multiple thresholds). For neural networks, we notice a performance drop when the waveform in the dataset changes substantially, or the critical frequency is not proper. New architectures or optimized hyper-parameter settings should be beneficial for neural networks to work well in these conditions. Besides, SAE does not display advantage over CRN in our simulations. This is preliminarily attributed to the model size (the amount of weights) of our neural networks. For efficient hardware implementation, the model size needs to be restricted. It is worthwhile to discuss effective measures to further improve the performance with limited weights in future researches.

\section{Conclusion}

In this paper, a simulation study on the waveform sampling--based detection procedure is conducted to evaluate the performance of traditional and emerging timing algorithms, including neural networks. Beyond the comparison between different algorithms, we introduce the estimation of Cram\'er Rao lower bound to provide more insights into the inner mechanism of feature extraction. In two case studies (single photon detection and shashlik-type calorimeter), we compare the lower bound with various algorithms for the prevalent and important task of pulse timing. These studies not only validate the methods of time measurement, but also demonstrate the practicality of the proposed lower bound in extensive application scenarios. We sincerely hope our work will provide a solid foundation for the research of feature extraction methods and benefit the development of radiation detectors in the future.

\appendix

\section{Cram\'er Rao lower bound for waveform sampling}
\label{sec:CRLB_for_waveform}

\subsection{Preliminary}

Let $\bm{R}$ be a vector of observations, and $\bm{\Theta}$ be a vector of random (or nonrandom) variables to be estimated. The conditional probability density function can be written as $p_{\bm{r}|\bm{\theta}}(\bm{R}|\bm{\Theta})$. Besides, for random variables, the prior probability density function can be written as $p_{\bm{\theta}}(\bm{\Theta})$.

According to the definition of conditional probability, the joint density function:

\begin{equation}
	p_{\bm{r},\bm{\theta}}(\bm{R},\bm{\Theta}) = p_{\bm{r}|\bm{\theta}}(\bm{R}|\bm{\Theta}) \cdot p_{\bm{\theta}}(\bm{\Theta})
\end{equation}

Thus:

\begin{equation}
	\ln p_{\bm{r},\bm{\theta}}(\bm{R},\bm{\Theta}) = \ln p_{\bm{r}|\bm{\theta}}(\bm{R}|\bm{\Theta}) + \ln p_{\bm{\theta}}(\bm{\Theta})
\end{equation}

In the signal estimation theory \cite{van2004detection}, the Fisher information matrix is used to compute the lower bound of estimation errors for a vector of variables. If we do not have any prior information, the matrix is determined by the probability conditioned on the estimated (nonrandom) variables. If the prior distribution is known, the posterior Cram\'er Rao Bound \cite{668800} can be applied and the matrix is determined by the joint distribution of observations and (random) variables.

For nonrandom variables, the Fisher information matrix is the expectation of the Hessian matrix of the conditional log-likelihood function on the variables:

\begin{equation} \label{equ:J_D}
	\bm{J}_D = -E \left[ \bm{\nabla}_{\bm{\Theta}} \left( \left\{ \bm{\nabla}_{\bm{\Theta}} \ln p_{\bm{r}|\bm{\theta}}(\bm{R}|\bm{\Theta}) \right\}^\mathrm{T} \right) \right]
\end{equation}

\begin{equation}
	\text{where   } {J_D}_{ij} = -E \left[ \frac{\partial^2 \ln p_{\bm{r}|\bm{\theta}}(\bm{R}|\bm{\Theta})}{\partial \Theta_i \partial \Theta_j} \right] \nonumber
\end{equation}

For random variables, the Fisher information matrix is comprised of two parts: information from observed data and information from prior distribution:

\begin{equation}
	\bm{J}_T \triangleq \bm{J}_D + \bm{J}_P
\end{equation}

\noindent where $\bm{J}_D$ is the same as above, and $\bm{J}_P$ is:

\begin{equation} \label{equ:J_P}
	\bm{J}_P = -E \left[ \bm{\nabla}_{\bm{\Theta}} \left( \left\{ \bm{\nabla}_{\bm{\Theta}} \ln p_{\bm{\theta}}(\bm{\Theta}) \right\}^\mathrm{T} \right) \right]
\end{equation}

\begin{equation}
	\text{where   } {J_P}_{ij} = -E \left[ \frac{\partial^2 \ln p_{\bm{\theta}}(\bm{\Theta})}{\partial \Theta_i \partial \Theta_j} \right] \nonumber
\end{equation}

\subsection{Modelling and computation of lower bound}
\label{sec:model_compute_lb}

We define the continuous signal waveform as:

\begin{equation}
	r(t) = s(t) + w(t),\quad w(t) \sim \mathcal{N}(0,\sigma_w^2)
\end{equation}

\noindent where $s(t)$ is the noiseless parameterized waveform function, and $w(t)$ is the noise term from a Gaussian random process. It should be noted that $w(t)$ is not necessarily independent between two time steps, because noise values at different sampling points can be correlated after passing through a linear time-invariant system. Here the quantization noise is seen as a contributing factor to $w(t)$ for simplicity, although it is discrete by nature.

We assume $s(t)$ is parameterized by $\bm{\Theta}$; furthermore, $\bm{\Theta}$ is divided into \emph{two domains}: $\bm{\Theta}_{\mathrm{int}}$ representing intrinsic parameters related to the detector response, and $\bm{\Theta}_{\mathrm{ext}}$ representing extrinsic parameters related to physical properties of the incident particle. Thus:

\begin{equation}
	s(t) \triangleq s(t;\bm{\Theta}) \triangleq s(t;\bm{\Theta}_{\mathrm{int}},\bm{\Theta}_{\mathrm{ext}})
\end{equation}

In later parts, if we do not make special explanations, the above three notations can be regarded as the same. The sampling process can be abstracted as recording the values of $r(t)$ from an origin at fixed intervals. The sampling points obey a multivariate Gaussian distribution:

\begin{equation} \label{equ:R_Theta}
	\bm{R|\Theta} = (R_1, R_2, ..., R_N|\bm{\Theta}) \sim \mathcal{N}(\bm{\mu}_{\bm{R}}, \bm{\Sigma}_{\bm{R}})
\end{equation}

\begin{equation}
	\text{where   } \bm{\mu}_{\bm{R}} = (s(0), s(\Delta), ..., s((N-1)\Delta)) \quad \text{(first sample at time 0)} \nonumber
\end{equation}

Using the distribution in equation (\ref{equ:R_Theta}) to compute the Fisher information matrix from data in equation (\ref{equ:J_D}), we can get:

\begin{align} \label{equ:J_D_signal}
	{J_D}_{ij} &= -E \left[ \frac{\partial^2}{\partial \Theta_i \partial \Theta_j} \ln \frac{\exp\left(-\frac 1 2 (\bm{R}-\bm{\mu}_{\bm{R}})^\mathrm{T}\bm{\Sigma}_{\bm{R}}^{-1}(\bm{R}-\bm{\mu}_{\bm{R}})\right)}{\sqrt{(2\pi)^N |\bm{\Sigma}_{\bm{R}}| } } \right] & \nonumber \\
	&= E \left[ \frac{\partial^2}{\partial \Theta_i \partial \Theta_j} \left( \frac 1 2 (\bm{R}-\bm{\mu}_{\bm{R}})^\mathrm{T}\bm{\Sigma}_{\bm{R}}^{-1}(\bm{R}-\bm{\mu}_{\bm{R}}) \right) \right] &\quad \text{(simplify and remove irrelevant terms)} \nonumber \\
	&= \left( \frac{\partial \bm{\mu}_{\bm{R}}}{\partial \Theta_i} \right)^{\mathrm{T}} \cdot \bm{\Sigma}_{\bm{R}}^{-1} \cdot \frac{\partial \bm{\mu}_{\bm{R}}}{\partial \Theta_j} &\quad \text{(take derivatives and expectation)}
\end{align}

The prior knowledge about the waveform is represented by the Fisher information matrix for prior distribution. To compute $\bm{J}_P$, we further assume $\bm{\Theta}_{\mathrm{int}}$ obey a multivariate Gaussian distribution with $\bm{\mu}_{\bm{\Theta}_{\mathrm{int}}}$ and $\bm{\Sigma}_{\bm{\Theta}_{\mathrm{int}}}$. In most cases, this is a reasonable assumption according to the central limit theorem, because many independent factors contribute to the detector response.

\begin{equation}
	\bm{\Theta}_{\mathrm{int}} \sim \mathcal{N}(\bm{\mu}_{\bm{\Theta}_{\mathrm{int}}}, \bm{\Sigma}_{\bm{\Theta}_{\mathrm{int}}})
\end{equation}

The only missing part for $\bm{J}_P$ and thus $\bm{J}_T$ is the distribution of $\bm{\Theta}_{\mathrm{ext}}$. This is very problem-dependent. For the timing problem discussed below, since the waveform sampling system and the incident particle are asynchronous by nature, a uniform distribution (with a relatively large range) fits the condition.

From the perspective of likelihood function, a uniform distribution will not affect the shape of the likelihood in its range, so it has no effect on the computed lower bound and thus provides no additional information. In view of this, we assume the inner elements of $\bm{\Theta}_{\mathrm{ext}}$ in $\bm{J}_P$ are zero.

Since $\bm{\Theta}_{\mathrm{int}}$ and $\bm{\Theta}_{\mathrm{ext}}$ are independent, we can get $\bm{J}_P$ as the following form through equation (\ref{equ:J_P}):

\begin{equation} \label{equ:J_P_signal}
	\bm{J}_P = \begin{pmatrix}
		\bm{\Sigma}_{\bm{\Theta}_{\mathrm{int}}}^{-1} & 0 \\
		0 & 0 \\
	\end{pmatrix}
\end{equation}

\noindent with first lines being from $\bm{\Theta}_{\mathrm{int}}$, and last lines from $\bm{\Theta}_{\mathrm{ext}}$.

Consider any estimator $\hat{\bm{\theta}}_i(\bm{R})$ of $\bm{\Theta}$. If we define the error vector $\bm{\theta}_{\epsilon}$ to be $\hat{\bm{\theta}}_i(\bm{R}) - \bm{\Theta}$, the correlation matrix of errors is:

\begin{equation}
	\bm{R}_{\epsilon} \triangleq E(\bm{\theta}_{\epsilon}\bm{\theta}_{\epsilon}^T)
\end{equation}

For diagonal elements, we have:

\begin{equation}
	E[{\theta_{\epsilon}}_i^2] \geq {J_T}^{ii}
\end{equation}

\noindent where ${J_T}^{ii}$ is the i-th diagonal element in the \emph{inverse} of the total information matrix. In other words, the diagonal elements of $\bm{J}_T^{-1}$ represent the minimum achievable variance when we estimate corresponding random variables and thus stand for the lower bound of any possible estimators.

In the above formulation, $\bm{J}_P$ serves as an important supplement to the total Fisher information matrix $\bm{J}_T$. Since $\bm{J}_P$ is singular, it cannot be used solely and must be combined with $\bm{J}_D$. In some cases when intrinsic parameters have fully correlated pairs, $\bm{J}_D$ can also become singular (as in section \ref{sec:cum_signal_sipm}). However, the sum $\bm{J}_T$ is always nonsingular and positive definite.

It should be noted that ${J_T}^{ii}$ is still a function of intrinsic and extrinsic parameters. The precise calculation needs to marginalize all parameters by Monte Carlo simulation. Here for simplicity, we use the most probable values of intrinsic parameters and sample extrinsic parameters from a uniform distribution. The result is averaged over all samples of extrinsic parameters.

\section{Case studies of the lower bound}
\label{sec:case_study_LB}

\subsection{Single photon signal of SiPM}
\label{sec:sp_signal_sipm}

Single photon detection is a common research topic in PET, and accurate timing is helpful to implement the so-called time-of-flight PET to reduce dosage and improve resolution. The single-cell current pulse is formulated as a function of circuit elements in the SiPM:

\begin{equation}
	s_{\mathrm{origin}}(t; \bm{\Theta}_{\mathrm{int}}, \bm{\Theta}_{\mathrm{ext}}) = f_{\mathrm{spad}}(t-\eta;C_d, V_{br}, R_q, C_q, C_g)
\end{equation}

\begin{equation}
	\text{where   } \bm{\Theta}_{\mathrm{int}} = \{ C_d, V_{br}, R_q, C_q, C_g \}, \quad \bm{\Theta}_{\mathrm{ext}} = \{ \eta \} \nonumber
\end{equation}

\noindent and $C_g$ is the lumped capacitance of the parasites between the anode and cathode of the SiPM. The response of the analog channel is characterized by its impulse response $h_{\mathrm{channel}}(t)$. Thus the signal at the waveform sampling side is the convolution between the original waveform and the channel response:

\begin{equation} \label{equ:s_t_single_photon}
	s(t; \bm{\Theta}) = \int_{-\infty}^{\infty} s_{\mathrm{origin}}(t-m; \bm{\Theta}) \cdot h_{\mathrm{channel}}(m) \mathop{}\!\mathrm{d} m
\end{equation}

Besides, we consider the electronic noise before and after the analog channel. An uncorrelated white Gaussian noise will become correlated after passing through the linear time-invariant system. If the sigma (standard deviation) of the original noise is $\sigma_d$, its autocorrelation at the waveform sampling side will be:

\begin{align}
	r_d(t) &= \sigma_d^2 \cdot (h_{\mathrm{channel}} \star h_{\mathrm{channel}})(t) & \nonumber \\
	&= \sigma_d^2 \cdot \int_{-\infty}^{\infty} h_{\mathrm{channel}}(t+m) \cdot h_{\mathrm{channel}}(m) \mathop{}\!\mathrm{d} m &\quad \text{(autocorrelation)}
\end{align}

By sampling $r_d(t)$ at fixed intervals, we can get the autocovariance matrix (because of zero mean) from the contribution of $\sigma_d$:

\begin{equation}
	\bm{\Sigma}_{\bm{Rd}} = \begin{pmatrix}
		r_d(0) & r_d(\Delta) & \dots & r_d((N-1)\Delta) \\
		r_d(\Delta) & r_d(0) & \dots & r_d((N-2)\Delta) \\
		\vdots & \vdots & \ddots & \vdots \\
		r_d((N-1)\Delta) & r_d((N-2)\Delta) & \dots & r_d(0)
	\end{pmatrix}
\end{equation}

Further assuming the sigma of the noise originated at the waveform sampling side is $\sigma_s$, the autocovariance matrix is diagonal because of the i.i.d condition:

\begin{equation}
	\bm{\Sigma}_{\bm{Rs}} = \sigma_s^2 \bm{I}
\end{equation}

Since noise originated from different sources is independent of each other, the total autocovariance matrix is the sum of above two:

\begin{equation} \label{equ:Sigma_single_photon}
	\bm{\Sigma}_{\bm{R}} = \bm{\Sigma}_{\bm{Rd}} + \bm{\Sigma}_{\bm{Rs}}
\end{equation}

With equation (\ref{equ:s_t_single_photon}) and equation (\ref{equ:Sigma_single_photon}), it is ready to compute the Fisher information matrix from data in equation (\ref{equ:J_D_signal}). Regarding the Fisher information matrix from prior distribution (equation (\ref{equ:J_P_signal})), it is reasonable to assume the variations of circuit elements are independent, so $\bm{\Sigma}_{\bm{\Theta}_{\mathrm{int}}}$ of the multivariate Gaussian distribution is a diagonal matrix.

\subsection{Cumulative signal of SiPM}
\label{sec:cum_signal_sipm}

For many applications other than PET, such as high energy physics, a single SiPM device receives hundreds of photons or more in a short period of time. Due to the non-trivial physical process, constructing a precise analytical function of the current pulse becomes intractable. Nevertheless, a semi-empirical function can serve as a reference and give directive results. As a fair approximation, we fit the cumulative signal of SiPM to the gamma distribution:

\begin{equation}
	f_{\mathrm{gamma}}(x, \alpha) = \frac{x^{\alpha - 1} e^{-x}}{\Gamma(\alpha)}
\end{equation}

\noindent where $x \geq 0$, $\alpha > 0$, and $\Gamma(\alpha)$ is the gamma function. The waveform is parameterized as:

\begin{equation}
	s_{\mathrm{origin}}(t; \bm{\Theta}_{\mathrm{int}}, \bm{\Theta}_{\mathrm{ext}}) = \kappa \cdot f_{\mathrm{gamma}}(\frac{t-\tau-\eta}{\beta}, \alpha) \label{signal_param}
\end{equation}

\begin{equation}
	\text{where   } \bm{\Theta}_{\mathrm{int}} = \{ \alpha, \tau, \beta, \kappa \}, \quad \bm{\Theta}_{\mathrm{ext}} =\{ \eta \} \nonumber
\end{equation}

For the electronic noise, since $\sigma_d$ is negligible compared to the large amplitude of the signal, the overall noise is dominated by the in-circuit electronic noise at the wave sampling side. Hence for simplification:

\begin{equation}
	\bm{\Sigma}_{\bm{R}} \approx \bm{\Sigma}_{\bm{Rs}} = \sigma_s^2 \bm{I} \approx \sigma_w^2 \bm{I}
\end{equation}

The Fisher information from the prior distribution is generated by a statistical method. We fit a sufficient amount of intrinsic parameters from raw data to a multivariate Gaussian distribution (with fixed $\eta$) and use the fitted covariance matrix as $\bm{\Sigma}_{\bm{\Theta}_{\mathrm{int}}}$. Other aspects of the modelling method are the same as section \ref{sec:sp_signal_sipm}.

\section{Details of the neural networks}

\subsection{Configuration}
\label{sec:NN_conf}

For all simulations, we use 8,000 examples for the training dataset and 2,000 examples for the test dataset. We choose Adam \cite{DBLP:journals/corr/KingmaB14} as the optimization algorithm. The initial learning rate is set to 0.001. The batch size is 64 when training, and we train for 30 epochs when the mean squared error (L2 loss) substantially decreases. We use Keras \cite{chollet2015} to implement the network model and follow all default configurations not discussed here. The model is deployed on a desktop computer with Intel(R) Core(TM) i7-10700F CPU, 32 GB RAM and RTX 2060 Super GPU (8 GB video memory).

\subsection{Network architecture}
\label{sec:NN_arch}

\begin{table}[H]
	\centering
	\scriptsize
	\begin{tabular}{cccccccc} 
		\hline
		layer name & input length & input channel & output length & output channel & kernel width & stride & activation \\
		\hline
		conv1 & 64 & 1 & 32 & 4 & 4 & 2 & ReLU \\
		conv2 & 32 & 4 & 16 & 8 & 4 & 2 & ReLU \\
		conv3 & 16 & 8 & 8 & 16 & 4 & 2 & ReLU \\
		conv4 & 8 & 16 & 4 & 32 & 4 & 2 & ReLU \\
		conv5 & 4 & 32 & 2 & 32 & 4 & 2 & none \\
		\hline
		deconv5 & 2 & 32 & 4 & 32 & 4 & 2 & ReLU \\
		deconv4 & 4 & 32 & 8 & 16 & 4 & 2 & ReLU \\
		deconv3 & 8 & 16 & 16 & 8 & 4 & 2 & ReLU \\
		deconv2 & 16 & 8 & 32 & 4 & 4 & 2 & ReLU \\
		deconv1 & 32 & 4 & 64 & 1 & 4 & 2 & none \\
		\hline
		fc1 & 64 (2*32) & -- & 64 & -- & -- & -- & ReLU \\
		fc2 & 64 & -- & 64 & -- & -- & -- & ReLU \\
		fc3 & 64 & -- & 1 & -- & -- & -- & none \\
		\hline
	\end{tabular}
	\caption{Hyper-parameters of neural networks with 64-point input. CRN is comprised of convolution and fully-connected layers, and SAE is comprised of convolution, deconvolution and fully-connected layers. ReLU standards for the rectified linear unit (an activation function to clip negative values to zero).}
	\label{tab:hyper_64_points}
\end{table}

\begin{table}[H]
	\centering
	\scriptsize
	\begin{tabular}{cccccccc} 
		\hline
		layer name & input length & input channel & output length & output channel & kernel width & stride & activation \\
		\hline
		conv1 & 16 & 1 & 8 & 8 & 4 & 2 & ReLU \\
		conv2 & 8 & 8 & 4 & 16 & 4 & 2 & ReLU \\
		conv3 & 4 & 16 & 2 & 32 & 4 & 2 & ReLU \\
		conv4 & 2 & 32 & 1 & 32 & 4 & 2 & none \\
		\hline
		deconv4 & 1 & 32 & 2 & 32 & 4 & 2 & ReLU \\
		deconv3 & 2 & 32 & 4 & 16 & 4 & 2 & ReLU \\
		deconv2 & 4 & 16 & 8 & 8 & 4 & 2 & ReLU \\
		deconv1 & 8 & 8 & 16 & 1 & 4 & 2 & none \\
		\hline
		fc1 & 32 (1*32) & -- & 64 & -- & -- & -- & ReLU \\
		fc2 & 64 & -- & 64 & -- & -- & -- & ReLU \\
		fc3 & 64 & -- & 1 & -- & -- & -- & none \\
		\hline
	\end{tabular}
	\caption{Hyper-parameters of neural networks with 16-point input. Refer to the description in table \ref{tab:hyper_64_points}.}
	\label{tab:hyper_16_points}
\end{table}

\acknowledgments

This research is supported by the National Key Research and Development Program of China (under project no. 2020YFE0202001). This research is also supported by China Postdoctoral Science Foundation (under grant no. 2021M690088).



\begin{thebibliography}{10}
	
	\bibitem{CERN-LHCC-2017-011}
	{\scshape CMS Collaboration}, \emph{{The Phase-2 Upgrade of the
			CMS Barrel Calorimeters}},  Tech. Rep. CERN-LHCC-2017-011. CMS-TDR-015, CERN,
	Geneva, Sep, 2017.
	
	\bibitem{Atanov:2018ich}
	N.~Atanov et~al., \emph{{The Mu2e Calorimeter Final Technical Design Report}},
	Tech. Rep., Feb, 2018.
	
	\bibitem{Semenov_2020}
	A.~Semenov, S.~Bazylev, E.~Belyaeva, M.~Bhattacharjee, B.~Dabrowska, D.~Egorov
	et~al., \emph{Electromagnetic calorimeter for {MPD} spectrometer at {NICA}
		collider},
	\href{https://doi.org/10.1088/1748-0221/15/05/c05017}{\emph{Journal of
			Instrumentation} {\bfseries 15} (2020) C05017}.
	
	\bibitem{Han_2019}
	Y.~Han, H.~Kang, S.~Song, G.~Ko, J.~Lee and S.~Hong, \emph{{SiPM}-based
		dual-ended-readout {DOI}-{TOF} {PET} module based on mean-time method},
	\href{https://doi.org/10.1088/1748-0221/14/02/p02023}{\emph{Journal of
			Instrumentation} {\bfseries 14} (2019) P02023}.
	
	\bibitem{PARK2019117}
	H.~Park and J.~S. Lee, \emph{Highly multiplexed SiPM signal readout for
		brain-dedicated {TOF}-{DOI} {PET} detectors},
	\href{https://doi.org/https://doi.org/10.1016/j.ejmp.2019.11.016}{\emph{Physica
			Medica} {\bfseries 68} (2019) 117}.
	
	\bibitem{FALLULABRUYERE2007247}
	A.~Fallu-Labruyere, H.~Tan, W.~Hennig and W.~Warburton, \emph{Time resolution
		studies using digital constant fraction discrimination},
	\href{https://doi.org/https://doi.org/10.1016/j.nima.2007.04.048}{\emph{Nuclear
			Instruments and Methods in Physics Research Section A: Accelerators,
			Spectrometers, Detectors and Associated Equipment} {\bfseries 579} (2007)
		247}.
	
	\bibitem{8694781}
	A.~{Shrestha} and A.~{Mahmood}, \emph{Review of Deep Learning Algorithms and Architectures}, \href{https://doi.org/10.1109/ACCESS.2019.2912200}{\emph{IEEE
			Access} {\bfseries 7} (2019) 53040}.
	
	\bibitem{doi:10.1146/annurev-nucl-101917-021019}
	D.~Guest, K.~Cranmer and D.~Whiteson, \emph{Deep Learning and Its Application to LHC Physics},
	\href{https://doi.org/10.1146/annurev-nucl-101917-021019}{\emph{Annual Review
			of Nuclear and Particle Science} {\bfseries 68} (2018) 161}.
	
	\bibitem{Ai_2019}
	P.~Ai, D.~Wang, G.~Huang, N.~Fang, D.~Xu and F.~Zhang, \emph{Timing and
		characterization of shaped pulses with {MHz} {ADCs} in a detector system: a
		comparative study and deep learning approach},
	\href{https://doi.org/10.1088/1748-0221/14/03/p03002}{\emph{Journal of
			Instrumentation} {\bfseries 14} (2019) P03002}.
	
	\bibitem{AI2020164420}
	P.~Ai, D.~Wang, G.~Huang, F.~Shen, N.~Fang, D.~Xu et~al., \emph{PulseDL: A
		reconfigurable deep learning array processor dedicated to pulse
		characterization for high energy physics detectors},
	\href{https://doi.org/https://doi.org/10.1016/j.nima.2020.164420}{\emph{Nuclear
			Instruments and Methods in Physics Research Section A: Accelerators,
			Spectrometers, Detectors and Associated Equipment} {\bfseries 978} (2020)
		164420}.
	
	\bibitem{Berg_2018}
	E.~Berg and S.~R. Cherry, \emph{Using convolutional neural networks to estimate
		time-of-flight from {PET} detector waveforms},
	\href{https://doi.org/10.1088/1361-6560/aa9dc5}{\emph{Physics in Medicine
			{\&} Biology} {\bfseries 63} (2018) 02LT01}.
	
	\bibitem{8844693}
	K.~{Gong}, E.~{Berg}, S.~R. {Cherry} and J.~{Qi}, \emph{Machine Learning in PET: From Photon Detection to Quantitative Image Reconstruction},
	\href{https://doi.org/10.1109/JPROC.2019.2936809}{\emph{Proceedings of the
			IEEE} {\bfseries 108} (2020) 51}.
	
	\bibitem{ZHOU2020787}
	D.-X. Zhou, \emph{Universality of deep convolutional neural networks},
	\href{https://doi.org/https://doi.org/10.1016/j.acha.2019.06.004}{\emph{Applied
			and Computational Harmonic Analysis} {\bfseries 48} (2020) 787}.
	
	\bibitem{10.1007/11840817_66}
	A.~M. Sch{\"a}fer and H.~G. Zimmermann, \emph{Recurrent Neural Networks Are Universal Approximators},  in \emph{Artificial Neural Networks -- ICANN
		2006}, S.~D. Kollias, A.~Stafylopatis, W.~Duch and E.~Oja, eds., (Berlin,
	Heidelberg), pp.~632--640, Springer Berlin Heidelberg, 2006.
	
	\bibitem{Belayneh:2019vyx}
	D.~Belayneh et~al., \emph{{Calorimetry with deep learning: particle simulation
			and reconstruction for collider physics}},
	\href{https://doi.org/10.1140/epjc/s10052-020-8251-9}{\emph{Eur. Phys. J. C}
		{\bfseries 80} (2020) 688}
	[\href{https://arxiv.org/abs/1912.06794}{{\ttfamily 1912.06794}}].
	
	\bibitem{Ai_2018}
	P.~Ai, D.~Wang, G.~Huang and X.~Sun, \emph{Three-dimensional convolutional
		neural networks for neutrinoless double-beta decay signal/background
		discrimination in high-pressure gaseous Time Projection Chamber},
	\href{https://doi.org/10.1088/1748-0221/13/08/p08015}{\emph{Journal of
			Instrumentation} {\bfseries 13} (2018) P08015}.
	
	\bibitem{PhysRevD.99.012011}
	P.~Baldi, J.~Bian, L.~Hertel and L.~Li, \emph{Improved energy reconstruction in
		NOvA with regression convolutional neural networks},
	\href{https://doi.org/10.1103/PhysRevD.99.012011}{\emph{Phys. Rev. D}
		{\bfseries 99} (2019) 012011}.
	
	\bibitem{AI2020164640}
	P.~Ai, D.~Wang, X.~Sun, G.~Huang and Z.~Li, \emph{A deep learning approach to
		multi-track location and orientation in gaseous drift chambers},
	\href{https://doi.org/https://doi.org/10.1016/j.nima.2020.164640}{\emph{Nuclear
			Instruments and Methods in Physics Research Section A: Accelerators,
			Spectrometers, Detectors and Associated Equipment} {\bfseries 984} (2020)
		164640}.
	
	\bibitem{Chen2020}
	J.-L. Chen, P.-C. Ai, D.~Wang, H.~Wang, N.~Fang, D.-L. Xu et~al., \emph{FPGA
		implementation of neural network accelerator for pulse information extraction
		in high energy physics},
	\href{https://doi.org/10.1007/s41365-020-00756-z}{\emph{Nuclear Science and
			Techniques} {\bfseries 31} (2020) 46}.
	
	\bibitem{CLINTHORNE1990157}
	N.~H. Clinthorne, A.~O. Hero, N.~A. Petrick and W.~Rogers, \emph{Lower bounds
		on scintillation detector timing performance},
	\href{https://doi.org/https://doi.org/10.1016/0168-9002(90)90767-Z}{\emph{Nuclear
			Instruments and Methods in Physics Research Section A: Accelerators,
			Spectrometers, Detectors and Associated Equipment} {\bfseries 299} (1990)
		157}.
	
	\bibitem{Seifert_2012}
	S.~Seifert, H.~T. van Dam and D.~R. Schaart, \emph{The lower bound on the
		timing resolution of scintillation detectors},
	\href{https://doi.org/10.1088/0031-9155/57/7/1797}{\emph{Physics in Medicine
			and Biology} {\bfseries 57} (2012) 1797}.
	
	\bibitem{Cates_2015}
	J.~W. Cates, R.~Vinke and C.~S. Levin, \emph{Analytical calculation of the
		lower bound on timing resolution for {PET} scintillation detectors comprising
		high-aspect-ratio crystal elements},
	\href{https://doi.org/10.1088/0031-9155/60/13/5141}{\emph{Physics in Medicine
			and Biology} {\bfseries 60} (2015) 5141}.
	
	\bibitem{GUNDACKER20156}
	S.~Gundacker, E.~Auffray, P.~Jarron, T.~Meyer and P.~Lecoq, \emph{On the
		comparison of analog and digital SiPM readout in terms of expected timing
		performance},
	\href{https://doi.org/https://doi.org/10.1016/j.nima.2014.10.020}{\emph{Nuclear
			Instruments and Methods in Physics Research Section A: Accelerators,
			Spectrometers, Detectors and Associated Equipment} {\bfseries 787} (2015) 6}.
	
	\bibitem{8049484}
	P.~{Lecoq}, \emph{Pushing the Limits in Time-of-Flight PET Imaging},
	\href{https://doi.org/10.1109/TRPMS.2017.2756674}{\emph{IEEE Transactions on
			Radiation and Plasma Medical Sciences} {\bfseries 1} (2017) 473}.
	
	\bibitem{Cramer:107581}
	H.~Cramer, \emph{{Mathematical methods of statistics}}, Princeton mathematical
	series. Princeton Univ. Press, Princeton, NJ, 1954.
	
	\bibitem{MENG2005435}
	L.~Meng and Z.~He, \emph{Exploring the limiting timing resolution for large
		volume CZT detectors with waveform analysis},
	\href{https://doi.org/https://doi.org/10.1016/j.nima.2005.04.076}{\emph{Nuclear
			Instruments and Methods in Physics Research Section A: Accelerators,
			Spectrometers, Detectors and Associated Equipment} {\bfseries 550} (2005)
		435}.
	
	\bibitem{AGOSTINELLI2003250}
	S.~Agostinelli et~al., \emph{Geant4—a simulation toolkit},
	\href{https://doi.org/https://doi.org/10.1016/S0168-9002(03)01368-8}{\emph{Nuclear
			Instruments and Methods in Physics Research Section A: Accelerators,
			Spectrometers, Detectors and Associated Equipment} {\bfseries 506} (2003)
		250}.
	
	\bibitem{NICA-MPD-ECAL-TDR}
	{MPD} {NICA} Technical Design Report of the Electromagnetic calorimeter (ECal)
	\url{http://mpd.jinr.ru/wp-content/uploads/2019/01/TDR_ECAL_v3.6_2019.pdf}.
	
	\bibitem{4179230}
	F.~{Corsi}, C.~{Marzocca}, A.~{Perrotta}, A.~{Dragone}, M.~{Foresta}, A.~{Del
		Guerra} et~al., \emph{Electrical Characterization of Silicon Photo-Multiplier Detectors for Optimal Front-End Design},  in \emph{2006 IEEE Nuclear Science
		Symposium Conference Record}, vol.~2, pp.~1276--1280, Oct, 2006,
	\href{https://doi.org/10.1109/NSSMIC.2006.356076}{DOI}.
	
	\bibitem{CORSI2007416}
	F.~Corsi, A.~Dragone, C.~Marzocca, A.~{Del Guerra}, P.~Delizia, N.~Dinu et~al.,
	\emph{Modelling a silicon photomultiplier (SiPM) as a signal source for optimum front-end design},
	\href{https://doi.org/https://doi.org/10.1016/j.nima.2006.10.219}{\emph{Nuclear
			Instruments and Methods in Physics Research Section A: Accelerators,
			Spectrometers, Detectors and Associated Equipment} {\bfseries 572} (2007)
		416}.
	
	\bibitem{5341428}
	S.~{Seifert}, H.~T. {van Dam}, J.~{Huizenga}, R.~{Vinke}, P.~{Dendooven},
	H.~{Lohner} et~al., \emph{Simulation of Silicon Photomultiplier Signals},
	\href{https://doi.org/10.1109/TNS.2009.2030728}{\emph{IEEE Transactions on
			Nuclear Science} {\bfseries 56} (2009) 3726}.
	
	\bibitem{ACERBI201916}
	F.~Acerbi and S.~Gundacker, \emph{Understanding and simulating SiPMs},
	\href{https://doi.org/https://doi.org/10.1016/j.nima.2018.11.118}{\emph{Nuclear
			Instruments and Methods in Physics Research Section A: Accelerators,
			Spectrometers, Detectors and Associated Equipment} {\bfseries 926} (2019)
		16}.
	
	\bibitem{Cova:96}
	S.~Cova, M.~Ghioni, A.~Lacaita, C.~Samori and F.~Zappa, \emph{Avalanche
		photodiodes and quenching circuits for single-photon detection},
	\href{https://doi.org/10.1364/AO.35.001956}{\emph{Appl. Opt.} {\bfseries 35}
		(1996) 1956}.
	
	\bibitem{NEURIPS2018_2a38a4a9}
	L.~Le, A.~Patterson and M.~White, \emph{Supervised autoencoders: Improving
		generalization performance with unsupervised regularizers},  in
	\emph{Advances in Neural Information Processing Systems}, S.~Bengio,
	H.~Wallach, H.~Larochelle, K.~Grauman, N.~Cesa-Bianchi and R.~Garnett, eds.,
	vol.~31, Curran Associates, Inc., 2018,
	
	\bibitem{8081803}
	W.~K. {Warburton} and W.~{Hennig}, \emph{New Algorithms for Improved Digital Pulse Arrival Timing With Sub-GSps ADCs},
	\href{https://doi.org/10.1109/TNS.2017.2766074}{\emph{IEEE Transactions on
			Nuclear Science} {\bfseries 64} (2017) 2938}.
	
	\bibitem{9187849}
	Y.~{Fan}, L.~{Zhao}, J.~{Qin}, Z.~{Jiang}, Z.~{Cao}, S.~{Liu} et~al.,
	\emph{Research and Verification on Real-Time Interpolated Timing Algorithm Based on Waveform Digitization},
	\href{https://doi.org/10.1109/TNS.2020.3022788}{\emph{IEEE Transactions on
			Nuclear Science} {\bfseries 67} (2020) 2246}.
	
	\bibitem{van2004detection}
	H.~L. Van~Trees, \emph{Detection, Estimation, and Modulation Theory, Part I:
		Detection, Estimation, and Linear Modulation Theory}. John Wiley \& Sons,
	2004.
	
	\bibitem{668800}
	P.~{Tichavsky}, C.~H. {Muravchik} and A.~{Nehorai}, \emph{Posterior Cramer-Rao bounds for discrete-time nonlinear filtering},
	\href{https://doi.org/10.1109/78.668800}{\emph{IEEE Transactions on Signal
			Processing} {\bfseries 46} (1998) 1386}.
	
	\bibitem{DBLP:journals/corr/KingmaB14}
	D.~P. Kingma and J.~Ba, \emph{Adam: {A} Method for Stochastic Optimization},
	in \emph{3rd International Conference on Learning Representations, {ICLR}
		2015, San Diego, CA, USA, May 7-9, 2015, Conference Track Proceedings},
	Y.~Bengio and Y.~LeCun, eds., 2015,
	\href{http://arxiv.org/abs/1412.6980}{http://arxiv.org/abs/1412.6980}.
	
	\bibitem{chollet2015}
	F.~Chollet, {Keras}. \url{https://github.com/keras-team/keras}, 2015.
	
\end{thebibliography}

\providecommand{\href}[2]{#2}\begingroup\raggedright\endgroup

\end{document}